\documentclass[12pt,preprint]{aastex}

\usepackage{epsfig}
\usepackage{lscape,graphicx,rotating}

\begin{document}

\title{Discovery of a Cosmological, Relativistic Outburst via its
  Rapidly Fading Optical Emission}

\author{S.~Bradley Cenko\altaffilmark{1},
  S.~R.~Kulkarni\altaffilmark{2},
  Assaf Horesh\altaffilmark{2},
  Alessandra Corsi\altaffilmark{3,4},
  Derek B.~Fox\altaffilmark{5},
  John Carpenter\altaffilmark{2},
  Dale A.~Frail\altaffilmark{6},
  Peter E.~Nugent\altaffilmark{7,1}
  Daniel A.~Perley\altaffilmark{2,8},
  D.~Gruber\altaffilmark{9},
  Avishay Gal-Yam\altaffilmark{10,11},
  Paul J.~Groot\altaffilmark{2,12},
  G.~Hallinan\altaffilmark{2},
  Eran O.~Ofek\altaffilmark{10},
  Arne Rau\altaffilmark{9},
  Chelsea L.~MacLeod\altaffilmark{13},
  Adam A.~Miller\altaffilmark{1},
  Joshua S.~Bloom\altaffilmark{1},
  Alexei V.~Filippenko\altaffilmark{1},
  Mansi M.~Kasliwal\altaffilmark{14},
  Nicholas M.~Law\altaffilmark{15},
  Adam N.~Morgan\altaffilmark{1},
  David Polishook\altaffilmark{16},
  Dovi Poznanski\altaffilmark{17},
  Robert M.~Quimby\altaffilmark{18},
  Branimir Sesar\altaffilmark{2},
  Ken J.~Shen\altaffilmark{7,1,19},
  Jeffrey M.~Silverman\altaffilmark{1,20},
  and Assaf Sternberg\altaffilmark{21}}

\altaffiltext{1}{Department of Astronomy, University of California,
   Berkeley, CA\,94720-3411, USA}
\altaffiltext{2}{Cahill Center for Astrophysics, California Institute of 
   Technology, Pasadena, CA 91125, USA}
\altaffiltext{3}{LIGO Laboratory, California Institute of Technology, MS 100-36,
    Pasadena, CA 91125, USA}
\altaffiltext{4}{Physics Department, George Washington University, 725
  21st St, NW Washington, DC 20052}
\altaffiltext{5}{Department of Astronomy \& Astrophysics, Pennsylvania State 
   University, University Park, PA 16802, USA}
\altaffiltext{6}{National Radio Astronomy Observatory, P.O. Box O, Socorro, 
   NM 87801, USA}
\altaffiltext{7}{Lawrence Berkeley National Laboratory, Berkeley, CA 94720, 
   USA}
\altaffiltext{8}{Hubble Fellow}
\altaffiltext{9}{Max Planck Institute for Extraterrestrial Physics,
    Giessenbachstrasse, Postfach 1312, 85748, Garching, Germany}
\altaffiltext{10}{Benoziyo Center for Astrophysics, Weizmann Institute of Science,
   76100 Rehovot, Israel}
\altaffiltext{11}{Kimmel Investigator}
\altaffiltext{12}{Department of Astrophysics/IMAPP, Radboud University Nijmegen, 
   6500 GL, Nijmegen, The Netherlands}
\altaffiltext{13}{Physics Department, United States Naval Academy, 572c 
   Holloway Roadd, Annapolis, MD 21402}
\altaffiltext{14}{Observatories of the Carnegie Institution for Science, 813
   Santa Barbara St., Pasadena, CA, 91101, USA}
\altaffiltext{15}{Dunlap Institute of Astronomy \& Astrophysics,
   University of Toronto, 50 St.~George Street, Toronto, ON M5S 3H4,
   Canada}
\altaffiltext{16}{Department of Earth, Atmospheric, and Planetary Sciences,
   Massachusetts Institute of Technology, Cambridge, MA 02139, USA}
\altaffiltext{17}{School of Physics and Astronomy, Tel-Aviv University, Tel Aviv
    69978, Israel}
\altaffiltext{18}{Kavli IPMU, University of Tokyo, 5-1-5 Kashiwanoha, Kashiwa
   City, Chiba 277-8583, Japan}
\altaffiltext{19}{Einstein Fellow}
\altaffiltext{20}{Department of Astronomy, University of Texas,
  Austin, TX 78712-0259, USA}
\altaffiltext{21}{Minerva Fellow, Max Planck Institute for Astrophysics, Karl 
   Schwarzschild St.~1, 85741 Garching, Germany}

\email{cenko@astro.berkeley.edu}


\shorttitle{\event: A Distant, Relativistic Outburst}
\shortauthors{Cenko \textit{et al.}}


\newcommand{\Swift}{\textit{Swift}}
\newcommand{\fermi}{\textit{Fermi}}
\newcommand{\mgii}{\ion{Mg}{2} $\lambda \lambda$ 2796, 2803}
\newcommand{\oii}{[\ion{O}{2} $\lambda$ 3727]}
\newcommand{\ip}{\textit{i$^{\prime}$}}
\newcommand{\zp}{\textit{z$^{\prime}$}}
\newcommand{\gp}{\textit{g$^{\prime}$}}
\newcommand{\rp}{\textit{r$^{\prime}$}}
\newcommand{\up}{\textit{u$^{\prime}$}}

\begin{abstract}
We report the discovery by the Palomar Transient Factory (PTF) of the
transient source PTF11agg, which is distinguished by three primary
characteristics: (1) bright ($R_{\mathrm{peak}} = 18.3$\,mag), rapidly
fading ($\Delta R =
4$\,mag in $\Delta t = 2$\,d) optical transient emission; (2) a faint
($R = 26.2 \pm 0.2$\,mag),
blue ($g^{\prime} - R = 0.17 \pm 0.29$\,mag) quiescent optical
counterpart; and (3) an associated year-long, scintillating radio
transient.  We argue that these observed properties are inconsistent
with any known class of Galactic transients (flare stars, X-ray
binaries, dwarf novae), and instead suggest a cosmological origin.  The
detection of incoherent radio emission at such distances implies a
large emitting region, from which we infer the presence of
relativistic ejecta.  The observed properties are all
consistent with the population of long-duration gamma-ray bursts
(GRBs), marking the first time such an outburst has been discovered in
the distant universe independent of a high-energy trigger.  We searched for
possible high-energy counterparts to PTF11agg, but found no evidence
for associated prompt emission.  We therefore consider three possible
scenarios to account for a GRB-like afterglow without a high-energy
counterpart: an ``untriggered'' GRB (lack of satellite coverage), an
``orphan'' afterglow (viewing-angle effects), and a ``dirty fireball''
(suppressed high-energy emission).  The observed optical and radio
light curves appear inconsistent with even the most basic predictions for 
off-axis afterglow models.  The simplest explanation, then, is that
PTF11agg is a normal, on-axis long-duration GRB for which the
associated high-energy emission was simply missed.  However, we have
calculated the likelihood of such a serendipitous discovery by PTF
and find that it is quite small ($\approx 2.6$\%).  While not definitive, we
nontheless speculate that PTF11agg may represent a new, more common
($> 4$ times the on-axis GRB rate at 90\% confidence) class of
relativistic outbursts lacking associated high-energy emission.  If
so, such sources will be uncovered in large numbers by future
wide-field optical and radio transient surveys.
\end{abstract}

\keywords{stars: flare -- stars: gamma-ray burst: general -- stars:
  supernovae}

\section{Introduction}
\label{sec:intro}
From accreting stellar-mass black holes in our Galaxy to distant
active galactic nuclei (AGNs) and gamma-ray bursts (GRBs), outflow
velocities approaching the speed of light are common in nature.
Indeed, the number of known sources capable of generating relativistic
ejecta has expanded in recent years to include a core-collapse
supernova without an accompanying GRB (SN\,2009bb; \citealt{scp+10}),
as well as the presumed tidal disruption of a star by a supermassive black hole
\citep{ltc+11,bgm+11,zbs+11,bkg+11,ckh+11}.  With revolutionary new
time-domain facilities slated to come online in the coming decade,
even more exotic examples will surely be uncovered.

Time-variable high-energy emission (X-rays and $\gamma$-rays) tends to be the
hallmark of such relativistic outflows.  Yet there is good reason to
expect that some relativistic outbursts may lack a detectable
high-energy signature.  In the case of GRBs, for example, the most
mundane possibility is a lack of sky coverage: the most sensitive
high-energy GRB detectors cover only a fraction of the sky at any
given time.  But other, more interesting
possibilities exist, including viewing-angle effects \citep{r97,pl98,npg02}
and some physical process suppressing the high-energy emission
entirely \citep{dcm00,hdl02,r03}.  The search at longer wavelengths 
for these ``orphan'' (i.e., off-axis) afterglows or ``dirty
fireballs'' has remained one of the most sought-after goals in the
GRB field for more than a decade.

In this work, we report the discovery by the Palomar Transient Factory
of PTF11agg, a rapidly fading optical transient associated with a 
year-long, scintillating radio counterpart.  The detection of a faint,
blue, quiescent optical source at the transient location suggests a
cosmological origin for the transient (i.e., well beyond the Milky Way
and any nearby galaxies).  At such distances, the observed radio
emission requires the presence of relativistic ejecta.  

Throughout this work, we adopt a standard $\Lambda$CDM cosmology with
H$_{0}$ = 71\,km s$^{-1}$ Mpc$^{-1}$, $\Omega_{\mathrm{m}} = 0.27$, and
$\Omega_{\Lambda} = 1 - \Omega_{\mathrm{m}} = 0.73$ \citep{sbd+07}.
 All quoted uncertainties are 1$\sigma$ (68\%) confidence
intervals unless otherwise noted, and UT times are used throughout.
Reported optical magnitudes are in the AB system \citep{og83}.  We
have corrected the reported optical and near-infrared (NIR) photometry
for a foreground Galactic extinction of $E(B-V) = 0.044$\,mag 
\citep{sfd98}, using the extinction law from \citet{ccm89}.

\section{Discovery and Basic Analysis}
\label{sec:obs}

\subsection{Optical/Near-Infrared}

\subsubsection{Observations}
\label{sec:opt}
Regular monitoring observations of field 100033 (centered at $\alpha =
08^{\mathrm{h}} 23^{\mathrm{m}} 32.42^{\mathrm{s}}$, $\delta =
+21^{\circ} 33\arcmin 34\farcs5$, with a total on-sky area of 7.2\,deg$^{2}$) 
were obtained with the Palomar 48\,inch Oschin telescope (P48) equipped 
with the refurbished CFHT12k camera \citep{rsv+08} as part of a program
to study stellar variability in Praesepe (the Beehive
Cluster; \citealt{acl+11}) by the Palomar Transient Factory 
(PTF; \citealt{lkd+09,rkl+09}).  Over 500 individual P48 frames, each with an
exposure time of 60\,s, were obtained over the period from 2009 
November through 2012 March. All P48 images were obtained with a 
Mould $R$-band filter, which is similar to the $r^{\prime}$ filter
from the Sloan Digital Sky Survey (SDSS; \citealt{aaa+11a}), but offset 
by $\sim 27$\,\AA\ redward \citep{oll+12}.  

\begin{figure}[t!]
  \plotone{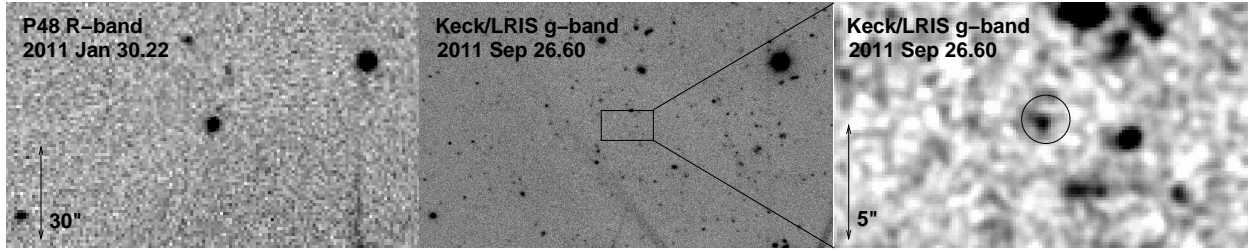}
 \caption[]{\footnotesize
  Optical imaging of the field of PTF11agg.  
  The P48 discovery
  ($R$-band) image is shown in the left panel.  Follow-up Keck/LRIS
  $g$-band observations, obtained on 2011 September 26, are 
  displayed in the center (wider field) and right (zoomed in) panels.  
  The location of PTF11agg, as determined from our P48
  imaging, is indicated with a solid circle (1\arcsec\ radius; note that
  this is significantly larger than the astrometric uncertainty in our
  alignment between the Keck/LRIS and P48 images, which is $\sim
  50$\,mas in each coordinate).  A faint, unresolved source
  consistent with the location of PTF11agg is
  detected in both our $g$-band and $R$-band (not shown) images.
  All images are oriented 
  with North facing up and East to the left.
 }
\label{fig:finder}
\end{figure}

In an image beginning at 5:17:11 on 2011 January 30, we detected a
bright but short-lived optical flare at the (J2000.0) location 
$\alpha = 08^{\mathrm{h}} 22^{\mathrm{m}} 17.195^{\mathrm{s}}$,
$\delta = +21^{\circ} 37\arcmin 38\farcs26$, with a 1$\sigma$ 
astrometric uncertainty of 70\,mas in each coordinate
(Figure~\ref{fig:finder}).  This source was subsequently dubbed 
PTF11agg by our automated discovery and classification pipeline 
\citep{brn+11}.  Our P48 photometry of PTF11agg, calculated with
respect to nearby point sources from SDSS, is presented in
Table~\ref{tab:opt}.  

\begin{figure}[t!]
  \plotone{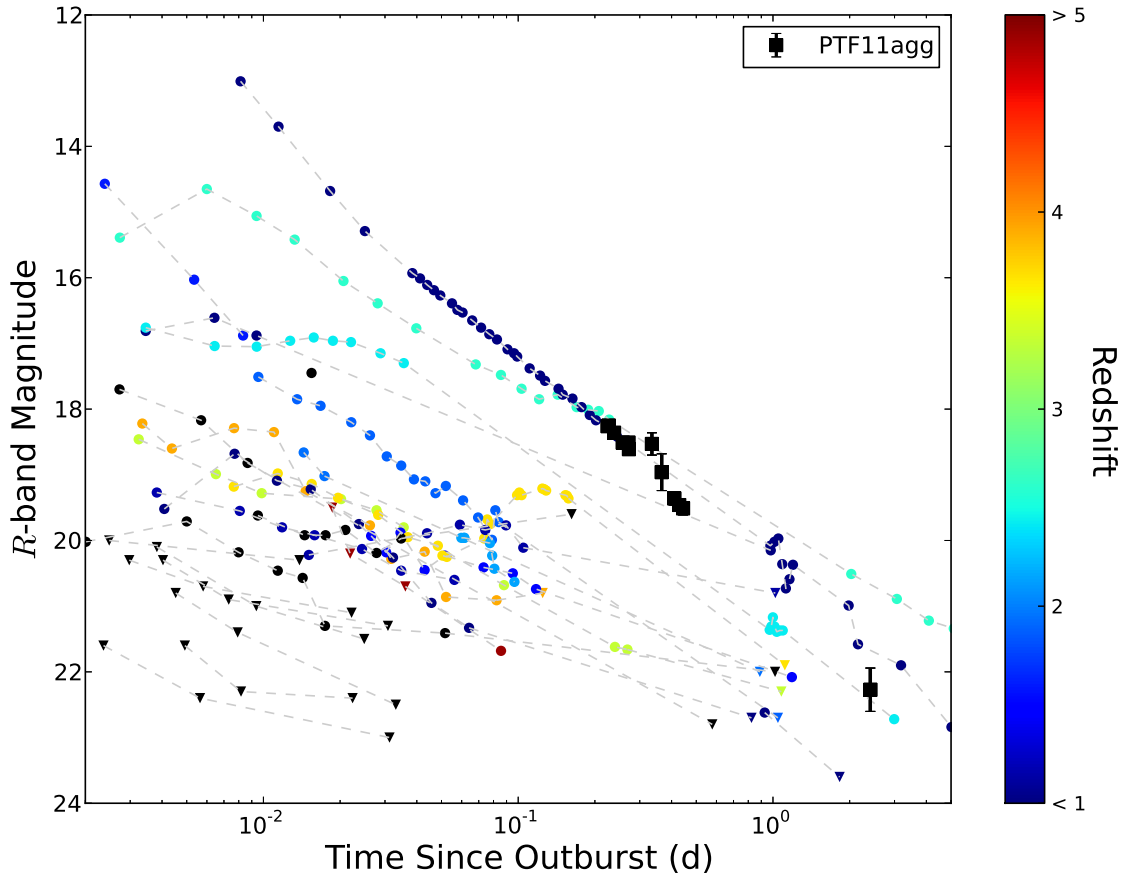}
  \caption[]{\footnotesize
  Optical light curve of PTF11agg, compared 
  with a representative
  sample of afterglows of long-duration GRBs discovered by the
  \textit{Swift} satellite \citep{ckh+09}.  The \textit{Swift} GRBs are
  color-coded by redshift; small black points indicate GRBs with unknown
  distance.  The observed power-law decline from PTF11agg ($\alpha 
  = 1.66$) is consistent with GRB afterglow observations at 
  $\Delta t \approx 1$\,d after the burst.  Though at the high end of the
  observed brightness distribution at $\Delta t \approx 0.2$\,d, a
  sizeable fraction ($\sim 10\%$) of \textit{Swift} events have a
  comparable $R$-band magnitude at $\Delta t \approx 1$\,d.  
  Inverted triangles mark 3$\sigma$ upper limits.
  }
\label{fig:olcurve}
\end{figure}

The peak observed magnitude, obtained in our first image of the field
on 2011 January 30, was measured to be $R = 18.26 \pm 0.05$\,mag.  
In the next ten P48 images of the field, all obtained on 2011 January
30, the source is seen to decay by 1.2\,mag in the $R$ band.  
A faint detection is also obtained by coadding all P48
images from 2011 February 1 ($R = 22.15 \pm 0.33$ mag).  
The resulting P48 $R$-band light curve is plotted in
Figure~\ref{fig:olcurve}.  All subsequent
P48 images result in nondetections at this location.

Examining our pre-outburst (i.e., before 2011 January 30) P48 imaging,
we find no evidence for emission at this location in any individual
frames (extending back in time to 2009 November).  The typical
limiting magnitude for an individual P48 image is $R \gtrsim 20$\,mag.
Stacking all frames from 2011 January 29 (i.e., the day preceding
discovery), we limit the optical emission at the location of PTF11agg
to $R > 21.9$\,mag.  Similarly, coadding all pre-outburst P48 images
results in a nondetection with $R > 23.7$\,mag.  

Deep optical imaging of the location of PTF11agg was obtained at late times
($\Delta t > 1$\,month) with the Low Resolution Imaging Spectrometer
(LRIS; \citealt{occ+95}) mounted on the 10\,m Keck I telescope
($g^{\prime}$- and $R$-band filters), and the Inamori-Magellan Areal
Camera and Spectrograph (IMACS; \citealt{dbh+11}) mounted on the 
6.5\,m Magellan-Baade telescope at Las Campanas Observatory 
($I$-band filter).

In our deepest epoch of post-outburst optical imaging (2011 
September 26 with Keck/LRIS, or $\Delta t = 240$\,d), we 
identify a faint, unresolved (in 0\farcs6 seeing) source in
$g^{\prime}$ and $R$ at (J2000.0) 
coordinates $\alpha = 08^{\mathrm{h}} 22^{\mathrm{m}}
17.202^{\mathrm{s}}$, $\delta = +21^{\circ} 37\arcmin 38\farcs26$
(Figure~\ref{fig:finder}).  
Given the uncertainty in the astrometric tie between the Keck/LRIS 
and P48 imaging (50 mas in each coordinate), the observed 90 mas 
radial offset is not statistically significant (null probability of 0.17).
Coadding Keck/LRIS images of the field of PTF11agg from several
individual nights with less ideal conditions (2011 March 4, March 12,
and April 27), we can recover an object at this location with similar
brightness in both $g^{\prime}$ and $R$.  No emission
is detected at this location in the $I$-band IMACS images to $I >
25.2$\,mag.

We obtained near-infrared (NIR) imaging of the location of
PTF11agg with the 1.3\,m Peters Automated InfraRed Imaging TELescope
(PAIRITEL; \citealt{bsb+06}) on 2011 March 1 ($\Delta t = 30$\,d).  A total 
exposure time of 2246\,s was obtained simultaneously in the $J$, $H$, 
and $K_{s}$ filters.  Raw data files were processed using standard NIR 
reduction methods via PAIRITEL Pipeline III (C.~Klein \textit{et al.}, in
preparation), and resampled using SWarp \citep{bmr+02} to create 1\farcs0
pixel$^{-1}$ images for final photometry. 

We also observed the location of PTF11agg with the Wide-Field Infrared
Camera (WIRC; \citealt{weh+03}) mounted on the 5\,m Hale telescope at
Palomar Observatory.  Images were obtained in the $K_{s}$ filter on
2012 March 28 ($\Delta t = 423$\,d) for a total exposure time of 
1200\,s.  The individual frames were reduced using a custom pipeline 
within the IRAF environment \citep{t86}.  Both the PAIRITEL and WIRC 
images were calibrated with respect to bright field stars from
the Two Micron All-Sky Survey (2MASS; \citealt{scs+06}).

No emission was detected at the location of PTF11agg in any of the NIR
bandpasses.  The most constraining limits come from the WIRC
observations ($K_{s} > 22.6$\,mag).

A full listing of our optical and NIR photometry is presented in
Table~\ref{tab:opt}.  To convert the Vega-based measurements from 2MASS 
to the AB system, we have used the offsets derived by \citet{br07}.  

\subsubsection{Constraints on the Decay Index and Optical Outburst Onset}
\label{sec:onset}
We fit the observed P48 detections on 2011 January 30 and February 1 to a
power-law model of the form $f_{\nu} = f_{0} (t - t_{0})^{-\alpha}$,
where $f_{\nu}$ is the flux density (in $\mu$Jy), $t_{0}$ is
the time of the outburst onset, $\alpha$ is the power-law index, and
$f_{0}$ is the flux density at a fiducial time ($t_{0} + 1$\,s).  We
find best-fit values of $\alpha = 1.66 \pm 0.35$ and $t_{0} = $ 23:34
UT ($\pm 1.7$\,hr) on 2011 January 29.  We note that the inferred
outburst onset $t_{0}$ occurs 16.6\,hr after the preceding P48
nondetection on 2011 January 29 (Table~\ref{tab:opt}).

In the event that PTF11agg is a \textit{bona fide} GRB-like afterglow
(\S\ref{sec:egal}), an alternate constraint on the explosion date can be derived by
comparing the peak brightness of PTF11agg with the
observed distribution of GRB optical afterglows.  Using
the comprehensive sample from \citet{kkz+10}, an
observed magnitude of $R = 18.26$ at discovery implies an age of
$\Delta t \lesssim 0.5$\,d (Figure~\ref{fig:olcurve}).  Put
differently, the brightest known GRB
optical afterglows reach an observed magnitude of $R \approx 18$
approximately 12 hours after the onset of the high-energy emission.
Together with the P48 nondetection on 2011 January 29.31, we can
conservatively constrain the outburst onset to fall within the window
from $\sim$ 17:00 on 2011 January 29 to 5:17 on 2011 January 30
(55590.71--55591.22 MJD).

While the overall power-law fit quality is acceptable ($\chi^{2} =
8.1$ for 9 degrees of freedom), we caution that the early optical
light curves of GRBs rarely exhibit single power-law
decays \citep{pv08,raa+09,ops+09}.  In the event that the outburst
occurred later than our derived $t_{0}$, the true power-law index will
be smaller than what we have inferred, and more consistent with most
previously observed GRB optical afterglows.  If the outburst actually
occurred earlier, the decay index would steepen somewhat.  But
temporal indices $\alpha \gtrsim 2.5$ are ruled out based on
the nondetection on 2011 January 29.  

\subsubsection{Likelihood of Quiescent Source Association}
\label{sec:pchance}
Here we wish to estimate $P_{\mathrm{chance}}$, 
the \textit{a posteriori} likelihood that the
coincident quiescent counterpart detected at late times in our
Keck/LRIS imaging is unrelated to PTF11agg (i.e., the transient
source).  We have measured the areal surface density of objects of
this brightness in our imaging of field 100033, finding $\sigma(R \leq 26.2)
 = 0.03$ galaxies arcsec$^{-2}$; we note that this is consistent
with the results from \citealt{hpm+97} using entirely different
fields.  Using 150\,mas, or three times
the uncertainty in the astrometric tie between the P48 and Keck/LRIS
images, as our search radius, and following \citet{bkd02}, we find
that $P_{\mathrm{chance}} = 2 \times 10^{-3}$.  We therefore consider it highly
likely that this source is the quiescent counterpart of PTF11agg;
however, we consider alternative possibilities below as well.

\subsection{Radio}

\subsubsection{Observations}
\label{sec:rad}
We began radio observations of the field of PTF11agg with
the National Radio Astronomy Observatory's (NRAO\footnote{The National
  Radio Astronomy Observatory is a facility of the National Science
  Foundation (NSF) operated under cooperative agreement by Associated
  Universities, Inc.})
Karl G.~Jansky Very Large Array (VLA; \citealt{pcb+11}) on 2011 March
11 ($\Delta t = 40$\,d).  The array was in the ``B'' configuration
until 2011 May 6, then the ``BnA'' configuration until 2011 June 1,
and the ``A'' configuration thereafter.  
Over the course of our monitoring, the angular
resolution ranged from 0\farcs3 to 1\farcs2.  The VLA data were 
reduced with the Astronomical Image Processing System 
(AIPS)\footnote{See http://www.aips.nrao.edu/index.shtml .}.  For flux 
calibration, we used the source 3C\,147, while phase calibration was
performed using the objects J0823+2223 and J0832+1832.  As a check of
our flux calibration, we have verified that flux measurements of our
phase calibration sources remain stable throughout the course of our
observations.  

We observed PTF11agg at high frequencies (mm wavelengths) with
the Combined Array for Research in Millimeter-wave Astronomy (CARMA)
beginning on 2011 March 14 ($\Delta t = 43$\,d) and continuing for 
approximately one month.
For our CARMA observations, the array was in the ``D'' configuration,
and the beam had an angular diameter of 10\arcsec.  The total 
bandwidth (lower sideband and upper sideband) was 8\,GHz, and the 
local oscillator frequency was 93.6\,GHz.  The
optical depth at high (230\,GHz) frequency ranged from fair ($\tau
\approx 0.4$; phase noise $\approx 50^{\circ}$) on March 14 and April
11, to good ($\tau \approx 0.1$; phase noise $\approx 40^{\circ}$) on
April 7.  Data were reduced using standard techniques within the 
MIRIAD environment \citep{stw95}.

\begin{figure}[t!]
  \plotone{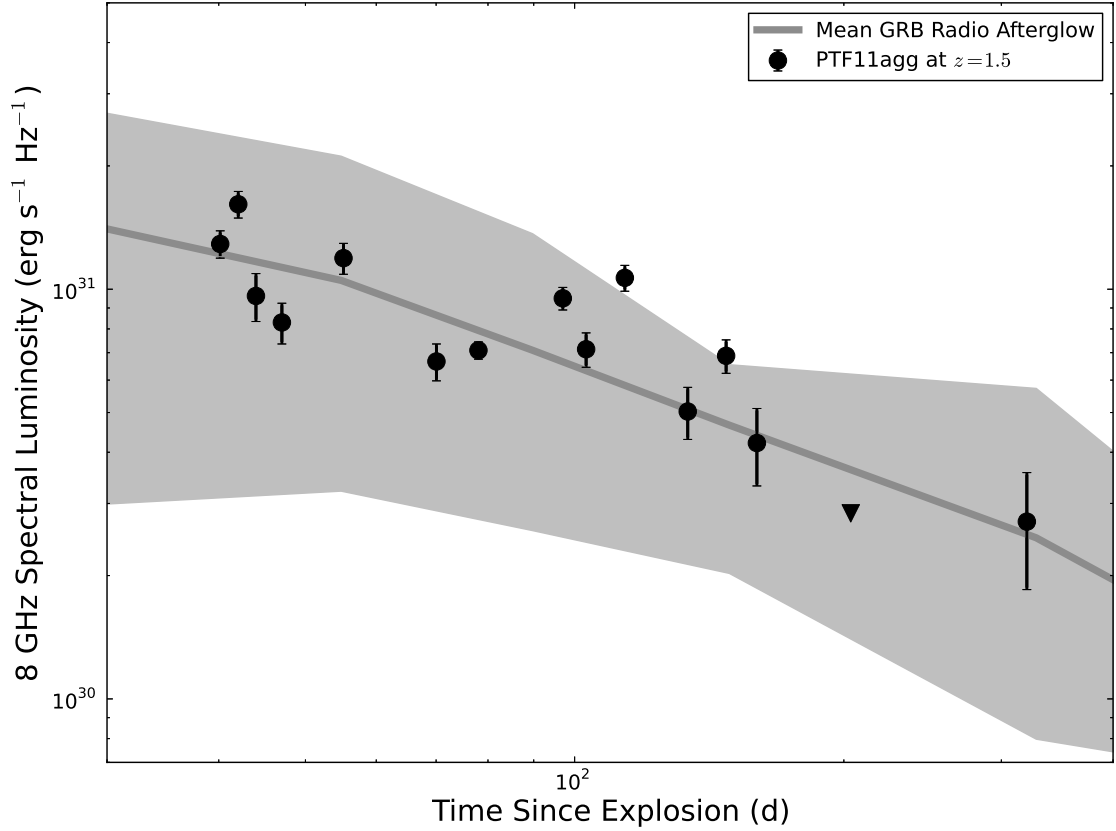}
  \caption[]{\footnotesize
  The 8\,GHz radio light curve of PTF11agg, 
  at an assumed redshift of
  1.5 (in the middle of our allowed range: $0.5 \lesssim z
  \lesssim 3.0$; \S\ref{sec:photoz}).  
  For comparison, we have plotted the mean
  long-duration GRB radio light curve (solid gray line), as well as
  the 25\%--75\% distribution (gray shaded region; \citealt{cf12}).  
  The variability
  superposed on the secular decline is likely due to interstellar
  scattering by electrons in the Milky Way, and is not intrinsic to the
  source.  For comparison, at $z = 0.5$, the 8\,GHz spectral
  luminosity would be a factor of 15 smaller, while at $z = 3.0$ a
  factor of 6 larger, than the values plotted here. 
  The inverted triangle marks a 3$\sigma$ upper limit.
  }
\label{fig:rlcurve}
\end{figure}

A transient radio counterpart was detected with both facilities.
The radio counterpart was unresolved (smallest beam size of 
$190$\,mas) and consistent with zero circular polarization 
($q \lesssim 10$\,\%) at all epochs.
The results of our EVLA and CARMA monitoring are displayed in
Table~\ref{tab:radio}, while the 8\,GHz light curve is plotted in
Figure~\ref{fig:rlcurve}.

\subsubsection{Spectral Energy Distribution}
\label{sec:radanalysis}
\begin{figure}[t!]
  \plotone{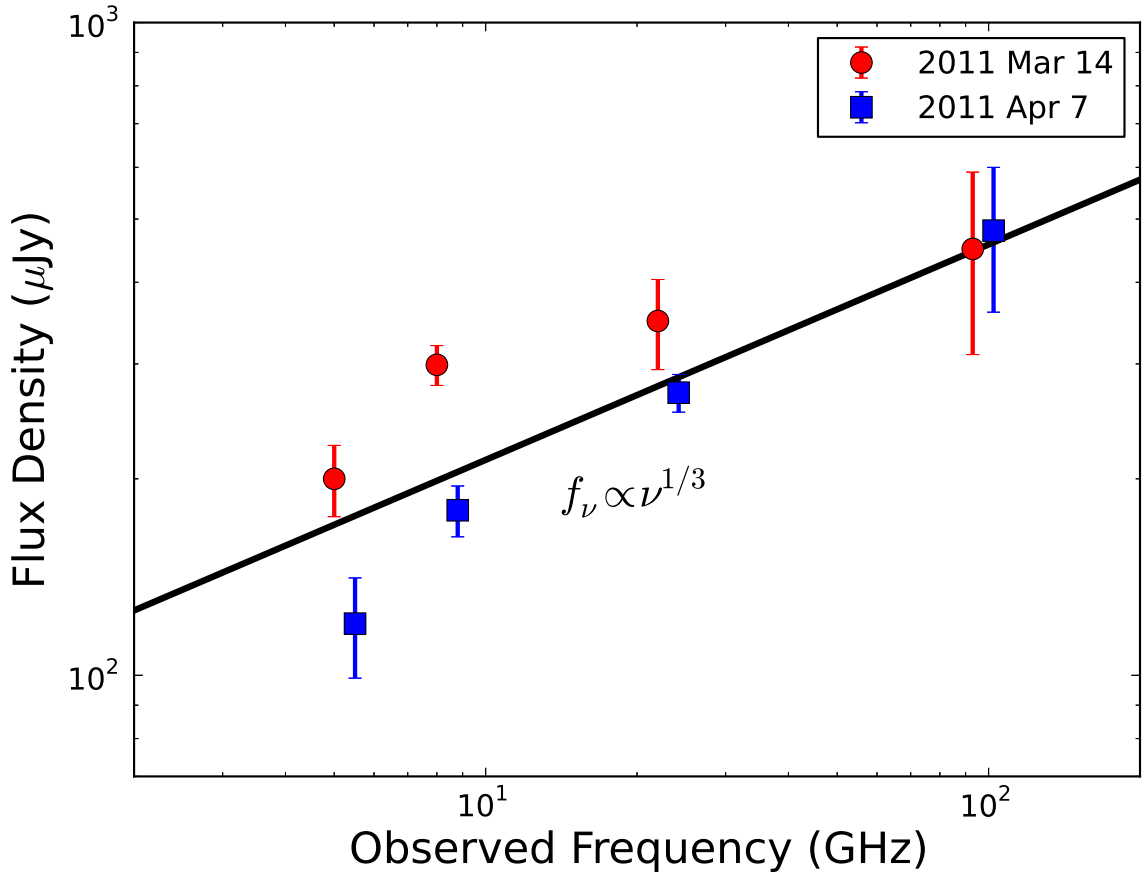}
  \caption[]{\footnotesize
  PTF11agg spectral energy distribution at radio frequencies.  The
  observations at lower frequencies have been interpolated to common
  epochs ($\Delta t \approx 43$ and 67\,d) to match the times of our
  CARMA observations.}
  \label{fig:radsed}
\end{figure}

To calculate the radio spectral energy distribution (SED), we must
interpolate the various observing frequencies to a common epoch.  To
provide the longest lever arm, we perform this analysis at the two
epochs of our 93\,GHz CARMA detections: 2011 March 14.05 ($\Delta t
\approx 43$\,d) and 2011 April 7.03 ($\Delta t \approx 67$\,d).  We
have linearly interpolated flux-density measurements made immediately
before and after these epochs at frequencies of 5 and 8\,GHz.  Due to
the relatively sparse coverage at 22\,GHz, we have simply adopted the
flux density at the closest epoch in time (note that for 2011 March 18
we averaged the two 22\,GHz measurements obtained on this day).  The
resulting SEDs are plotted in Figure~\ref{fig:radsed}.

We fit a power law of the form $f_{\nu} = f_{0} \nu^{\beta}$ to the
data, where $f_{\nu}$ is the flux density (in $\mu$Jy), $\nu$ is the
observing frequency (in GHz), $\beta$ is the power-law spectral index,
and $f_{0}$ is the flux density at a fiducial frequency of 1\,GHz.
For the first epoch ($\Delta t \approx 43$\,d), we find $\beta = 0.28
\pm 0.08$.  On the second epoch ($\Delta t \approx 67$\,d), we measure
$\beta = 0.46 \pm 0.07$.  Given the relatively large degree of
variability (see below), together with the sparse coverage at high
frequencies, we adopt $\beta = 1/3$ as an approximate
spectral slope in the radio for the remainder of this work.

\subsubsection{Angular Source Size}
\label{sec:size}
The presence of nonthermal radio emission provides two powerful and
independent means to constrain the angular size of the emitting region.
To begin with, the brightness temperature ($T_{B}$)
of an incoherent radio emitter cannot exceed its equipartition value of
$T_{B,\mathrm{eq}} \approx 10^{11}$\,K \citep{r94,kfw+98}.  
The brightness temperature is given by
\begin{equation}
T_{B} = \frac{c^{2}}{2 k_{B} \nu^{2}} \frac{f_{\nu}}{\pi \Theta^{2}},
\label{eqn:tb}
\end{equation}
where $c$ is the speed of light, $k_{B}$ is Boltzmann's constant,
$\nu$ is the observing frequency, $f_{\nu}$ is the observed flux
density, and $\Theta$ is the angular diameter of the emitting region.
Adopting $T_{B} \lesssim 10^{11}$\,K in Equation~\ref{eqn:tb} thus
implies a lower limit on the angular diameter of the source:
\begin{equation}
\Theta \gtrsim 2.1\,\left( \frac{f_{\nu}}{\mu\mathrm{Jy}}
  \right)^{1/2}\,\left( \frac{\nu}{\mathrm{GHz}}
  \right)^{-1}\,\,\mu\mathrm{as}.
\label{eqn:size1}
\end{equation}

As can be seen from Equation~\ref{eqn:size1}, the strictest lower
limits on the size of the emitting region are derived from
observations at the lowest frequencies (assuming a power-law spectral
index $\beta < 2$).  Using our 4.5\,GHz observation on 2011 March 13,
we find $\Theta > 7$\,$\mu$as.  Most of our early observations at 5
and 8\,GHz yield comparable (though slightly less strict) limits.  
 
Separately, we can constrain the angular size of the source from the
detection of interstellar scattering and scintillation (ISS; 
\citealt{r90}).  To
quantify the degree of variation induced by the scattering electrons,
we calculate the modulation index,
\begin{equation}
m_{p}(\nu) = \frac{ \sqrt{ V(f_{\nu}) - \langle \sigma^{2} \rangle } }
    {\langle f_{\nu} \rangle},
\label{eqn:modindex}
\end{equation}
where $V(f_{\nu})$ is the variance of the flux density (with respect to
an assumed model), $\langle \sigma^{2} \rangle$ is the average of the
square of the individual measurement uncertainties, and $\langle f_{\nu}
\rangle$ is the average of the flux density.

We calculated the modulation indices at 5 and 8\,GHz, neglecting higher
frequencies due to the relative lack of observations.  We fit the light
curves at both frequencies to a power-law model of the form $f_{\nu} =
f_{0} (t - t_{0})^{-\alpha}$, finding best-fit temporal indices of $\alpha_{5\,\mathrm{GHz}} =
-0.09 \pm 0.13$ (i.e., consistent with no temporal evolution) and
$\alpha_{8\,\mathrm{GHz}} = 0.56 \pm 0.06$.  This power-law model then forms the
reference which we use to calculate the variance at each frequency.
In this manner, we find $m_{p}(5$\,GHz$) = 0.42$ and $m_{p}(8$\,GHz$)
= 0.26$.

We use the Galactic electron density distribution model of
\citet{cl02} to derive the relevant ISS parameters, namely
$\nu_{0}$, the transition frequency between the strong and weak 
scattering regimes.  For the line of sight to PTF11agg (Galactic
coordinates $l = 202.08^{\circ}$, $b = 29.2^{\circ}$), we find
$\nu_{0}  = 11$\,GHz.  For a point source, the maximum degree of
modulation ($m_{p} = 1$) will occur at this transition frequency.  It
is therefore not unreasonable to expect our observations at 5, 8, and
(possibly) 22\,GHz to suffer from some degree of ISS.

For $\nu_{0} = 11$\,GHz, our observations at 5 and 8\,GHz will be in
the strong scattering regime ($\nu < \nu_{0}$).  Furthermore, given
the relatively broad bandwidth of our observations ($\Delta \nu / \nu 
\approx 0.1$), we consider only refractive scintillation.
For a point source, the modulation index in the strong, refractive
regime is given by \citep{w98}
\begin{equation}
m_{p}(\nu) = \left( \frac{\nu}{\nu_{0}} \right)^{17 / 30}.
\label{eqn:iss1}
\end{equation}
For the line of sight to PTF11agg, we therefore expect a significant
degree of modulation for a point source at our observing frequencies:
$m_{p}(5$\,GHz$) = 0.63$, $m_{p}(8$\,GHz$) = 0.82$.  

For an extended source, the observed modulation will be reduced by a
factor of $(\Theta_{r} / \Theta)^{7/6}$, where $\Theta_{r}$ is
the size of the Fresnel scattering disk \citep{w98},
\begin{equation}
\Theta_{r} = \frac{8}{\sqrt{D \nu_{0}}} \left( \frac{\nu_{0}}{\nu}
\right)^{11/5}\,\mu\mathrm{as},
\label{eqn:iss2}
\end{equation}
where $D$ is the effective distance to the scattering screen ($D =
0.78$\,kpc for the line of sight to PTF11agg).  If we solve for the
angular diameter corresponding to the observed degree of modulation at
each frequency, we find $\Theta(8$\,GHz$) = 10$\,$\mu$as and
$\Theta(5$\,GHz$) = 34$\,$\mu$as.  We therefore conclude that the
angular size of the emitting region at $\Delta t \approx 100$\,d is
$\Theta \approx 20$\,$\mu$as.  

\subsection{High-Energy}

\subsubsection{$\gamma$-ray Limits}
\label{sec:gamma}
At the time of discovery, three primary high-energy facilities
were monitoring the sky to search for the prompt emission from GRBs.  
The Third InterPlanetary Network (IPN; \citealt{hga+10}) is a group of 
nine satellites sensitive to high-energy emission.  When multiple
satellites detect a GRB, the sky localization can be reconstructed 
from light travel time constraints.  The IPN provides essentially 
continuous all-sky coverage (i.e., 100\% duty cycle), with a
sensitivity to fluences (10\,keV -- 5\,MeV) of $S_{\gamma} \gtrsim 6 
\times 10^{-7}$\,erg\,cm$^{-2}$ (at 50\% efficiency; i.e., half of the
GRBs with this fluence are too faint to trigger the IPN detectors).  
In addition to the IPN, the Gamma-Ray Burst Monitor (GBM; \citealt{mlb+09}) on
the \textit{Fermi} satellite, and the Burst Alert Telescope
(BAT; \citealt{bbc+05}) on the \textit{Swift} satellite \citep{gcg+04}, 
also regularly discover a large number of GRBs.  The GBM detects 
bursts down to a 8\,keV -- 1\,MeV fluence of $S_{\gamma} \gtrsim 
4 \times 10^{-8}$\,erg\,cm$^{-2}$, but has a field of view of
8.8\,sr (the area of the sky unocculted by the Earth in the
\textit{Fermi} orbit) and a duty cycle of $\gtrsim 80$\%.  Likewise,
the \textit{Swift} BAT has detected events with 15--150\,keV fluences
as low as $6 \times 10^{-9}$\,erg\,cm$^{-2}$, but only observes a
field of view of 2\,sr with a duty cycle of $\sim 90\%$.  We
caution that, for all three facilities, the high-energy fluence
required to trigger the onboard GRB algorithms depends on the duration
of the event; therefore, the above sensitivity limits should be
treated only as approximate.
 
We have
searched all three facilities for GRB triggers from the direction of
PTF11agg over the time period from 17:00 2011 January 29 
to 5:17 2011 January 30 (i.e., the outburst onset window derived in
\S\ref{sec:onset}).  No triggers
were reported by any facility in the direction of PTF11agg during this
$\sim 12$\,hr window.  We further conducted a search for untriggered events
in the GBM data in the energy range 10--300\,keV on several
different time scales (0.256\,s, 0.512\,s, 1.024\,s, 2.048\,s,
4.096\,s, and 8.192\,s)\footnote{Two individual GBM detectors with a
  significance of 4.0$\sigma$ and 3.8$\sigma$ above background were required
  for a trigger to register in this search.}.  No potential
high-energy counterparts to PTF11agg were found.    

Given the field of view and duty cycle of the GBM and BAT, there is a
significant likelihood that events below the IPN sensitivity
threshold would be missed by both instruments.  For example, for a GRB
with fluence above the GBM sensitivity level (but below the IPN
threshold), the probability of a nondetection from \textit{both} 
instruments is as high as $\sim 40\%$ (assuming a uniform and 
independent distribution of sky pointings for the two instruments).  
Given the relatively large window of time required to search (i.e.,
multiple \textit{Swift} and \textit{Fermi} orbits), we 
consider a fluence of $S_{\gamma} \lesssim 10^{-6}$\,erg\,cm$^{-2}$ 
(i.e., twice the all-sky IPN sensitivity) a reasonable limit on
any high-energy prompt emission associated with PTF11agg.  Given the
extremely weak correlation between prompt $\gamma$-ray fluence 
and optical afterglow brightness \citep{nfp09}, this limit is consistent 
with the known properties of GRBs and their afterglows.

\subsubsection{X-ray Limits}
\label{sec:xray}
To search for an X-ray counterpart, we obtained observations of the 
location of PTF11agg with the X-ray Telescope (XRT; \citealt{bhn+05}) on 
board the \textit{Swift} satellite on 2011 March 13 ($\Delta t = 42$\,d).  
Data were reduced using the automated pipeline described by 
\citet{bk07}.  No X-ray source is detected at the location of 
PTF11agg at this time.  Assuming a power-law spectrum with a 
photon index of $\Gamma = 2$, we derive a $3\sigma$ upper limit 
on the 0.3--10\,keV flux of $f_{X} < 8 \times 
10^{-14}$\,erg\,cm$^{-2}$\,s$^{-1}$.  

Finally, we note that no historical X-ray emission has been reported
at this location, neither in the {\it ROSAT} All-Sky Survey
(0.1--2.4\,keV; \citealt{vab+99}) nor in any compilations of known Galactic
X-ray sources (accessed via the HEASARC\footnote{See 
http://heasarc.gsfc.nasa.gov .} and SIMBAD\footnote{See
http://simbad.u-strasbg.fr/simbad .} databases).

\section{Comparison with Known Galactic Transients}
\label{sec:galactic}
The combination of (1) a rapidly fading optical transient ($\Delta R
\gtrsim 4$\,mag in $\Delta t = 2$\,d) and (2) a faint, blue ($g^{\prime} -
R = 0.17 \pm 0.29$\,mag), quiescent optical counterpart makes PTF11agg
unique amongst the thousands of discoveries by PTF to date. 
Together with (3) the long-lived ($\Delta t \approx 300$\,d) radio 
emission, here we attempt to simultaneously account for these three 
distinguishing characteristics.

In order to understand the nature of the emission from PTF11agg, 
we must constrain its distance.  In this section, we first consider a
Galactic origin by comparing PTF11agg with known classes of Galactic
transients.

Assuming the faint optical source is associated with PTF11agg (i.e.,
the quiescent counterpart), the measured color, $g^{\prime} - R
= 0.17 \pm 0.29$\,mag, implies a spectral type of
$\sim$ F2 ($T_{\mathrm{eff}} \approx 7000$\,K) for a main-sequence
star.  More conservatively, adopting our 3$\sigma$ limit on the color
($g^{\prime} - R < 1.04$\,mag), we can rule out single main-sequence stars with
$T_{\mathrm{eff}} \lesssim 4500$\,K (i.e., cooler than spectral type
K4).  Given the observed brightness, a main-sequence star hotter than
K4 would lie at a distance $d \gtrsim 90$\,kpc.  This firmly rules out an
association with the Praesepe cluster ($d \approx 175$\,pc); in fact,
only 4 globular clusters are known to exist at such large distances in
the extreme outer halo of the Milky Way (e.g., AM\,1 at $d \approx
120$\,kpc; \citealt{ma79}).  In addition to the extremely small source
densities this far in the halo, the inferred lower limit on the radio
luminosity at such a distance ($\nu L_{\nu} \gtrsim
10^{31}$\,erg\,s$^{-1}$) is three orders of magnitude larger than the
most luminous known stellar radio sources (e.g., RS CVn binaries, FK
Com class stars, and Algol-class stars; \citealt{g02}).

While \textit{a posteriori} unlikely, it is nonetheless important to
consider that the quiescent optical source may be unrelated to 
PTF11agg.  Absent color information, an optical nondetection, even at
the depth of our late-time imaging, is not sufficient to rule out a
Galactic origin.  With their smaller effective temperatures, low-mass
stars and brown dwarfs (in particular ultracool stars, with spectral
type later than M7) emit little flux in the optical bandpass.
Furthermore, ultracool stars are known to exhibit high-amplitude, short
timescale (minutes to hours) optical and radio outbursts that have in
the past been mistaken for extragalactic transients 
\citep{bwb+04,kr06,ATEL.4586,ATEL.4619}.

We can use our NIR limits on the quiescent emission at the location of
PTF11agg to calculate the minimum distance to
an ultracool star as a function of spectral type; in other words, for
each spectral type, any object closer than this ``detectability''
distance would be identified in our NIR imaging.  For spectral types
later than $\sim$ M4, the strongest constraint is provided by our
deepest epoch of $K_{s}$-band imaging: $K_{s} > 22.6$\,mag.  Using the 
observed $K_{s}$-band magnitudes and distance (parallax) measurements for
ultracool stars from \citet{dhv+02} and \citet{psb+06}, we fit a low-order polynomial to
calculate the absolute $K_{s}$-band magnitude
as a function of spectral type, $M_{K_{s}} (ST)$, where $ST = 5$ for
M5, $ST = 12$ for L2, etc.  We find the scatter about our derived
absolute $K_{s}$-band magnitude fit is $\sim 0.30$\,mag (i.e., 30\%)
over the range M5--T8.  We then convert the observed peak 
radio flux density ($f_{\nu,\mathrm{peak}} \approx 300$\,$\mu$Jy) to 
a lower limit on the radio luminosity ($\nu L_{\nu}$) using these
distance constraints.  

\begin{figure}[t!]
  \plotone{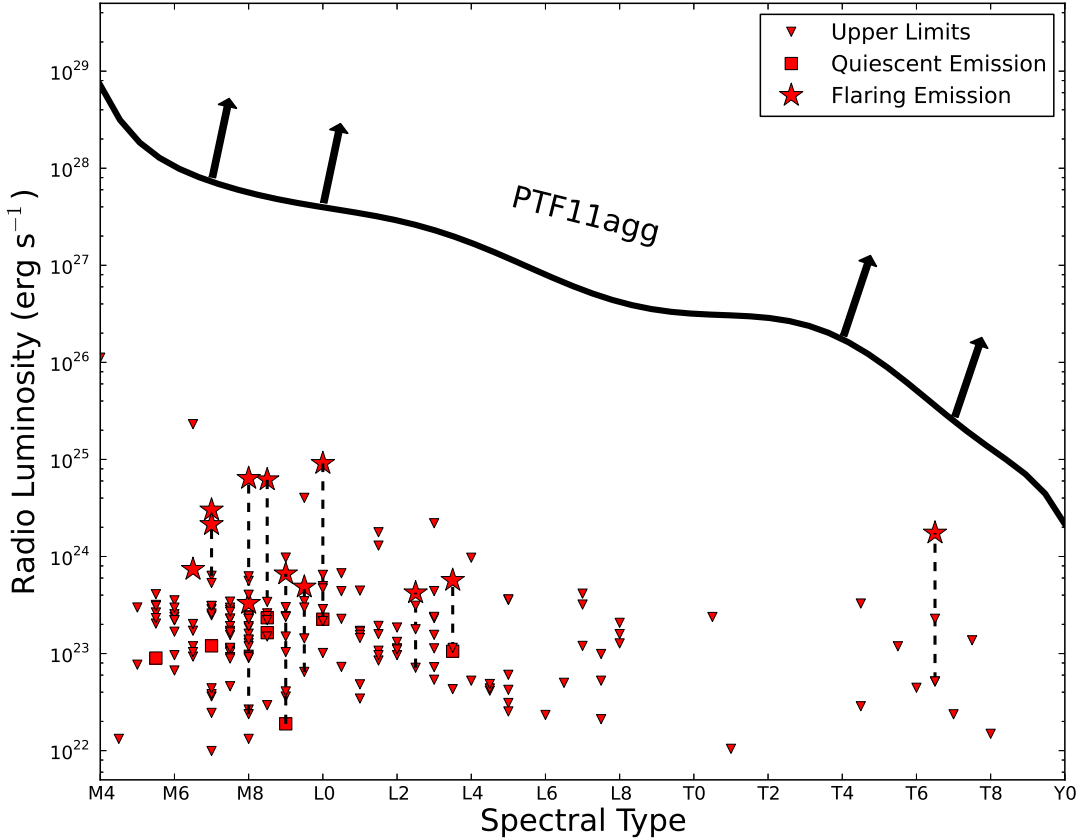}
  \caption[]{\footnotesize
   Lower limits on the radio luminosity of PTF11agg (solid black
   line).  For each spectral type, we calculate a minimum 
  ``detectability'' distance using limits from our 
  NIR imaging (i.e., any source more nearby would have been detected).  
  We then convert this distance to a lower limit on the radio
  luminosity based on the observed peak flux from PTF11agg.  Shown 
  for comparison are radio observations of ultracool stars from the
  literature \citep{bbf+10,b06,adh+07,aob+07,b02,bbb+01,brp+09,bbg+08,bgg+08,brr+05,bp05,hbl+07,rw12,mbr12}.
  The inferred luminosity is several orders of magnitude larger than that of 
  any previously observed low-mass star or brown dwarf, either in a 
  quiescent or flaring state.
}
\label{fig:mdwarfs}
\end{figure}

The resulting luminosity limits, as a function of spectral type, 
are plotted in Figure~\ref{fig:mdwarfs}.
For comparison, we have also plotted all
radio observations of ultracool stars from the literature (see figure
caption for references).  Our luminosity limits are typically at least
two orders of magnitude larger than the most luminous known ultracool
stellar flares.  Even comparing with the recently detected flare from
the T6.5 dwarf 2MASS J1047+21, by far the coolest brown dwarf
detected at radio frequencies \citep{rw12}, our limits require a radio
luminosity a factor of $> 20$ times larger.  We furthermore 
see no evidence for a high degree of circular polarization (common 
to many, though not all, flares; \citealt{b06,hbl+07}), and the radio emission from 
PTF11agg is much more long-lived than these low-mass stellar outbursts 
(durations typically of only hours).  

In addition to stellar flares, binary systems where one member is a
compact object (white dwarf, neutron star, or black hole) are known
sources of optical and radio outbursts in the Milky Way.  Such a
system could circumvent two issues with Galactic transients we
previously identified.  First, the energy release during the accretion
process is more than sufficient to power the observed radio flux; Cyg
X-3 \citep{gjs+83}, for example, has reached peak radio luminosities 
in excess of $10^{34}$\,erg\,s$^{-1}$.  Second, the presence of an 
accretion disc can alter the optical color of such
systems.  Accordingly, our previous inference that the quiescent 
counterpart must lie at $d \gtrsim 90$\,kpc would no longer be 
valid.

We consider first X-ray binaries, where the degenerate primary is a
neutron star or black hole.  Of particular interest are the subclass 
of microquasars \citep{mr99}, whose powerful radio jets exhibit 
apparent superluminal motion (and thus imply a relativistic outflow).
Due to the lack of a bright quiescent optical counterpart, we consider
only low-mass systems where the accretion occurs via Roche-lobe
overflow from the nondegenerate secondary (low-mass X-ray binaries,
or LMXBs).

Black hole LMXBs are typically characterized by
well-defined ``states'': correlations between X-ray spectra,
X-ray flux, and radio emission \citep{rm06}.  Radio emission is observed
in a well-defined region of this hardness-intensity phase
space \citep{fbg04,fkm04}.  In the so-called low-hard state, thought to
correspond to low ($\lesssim 0.01 L_{\mathrm{Edd}}$), radiatively
inefficient accretion \citep{emn97}, relatively steady radio emission 
from a jet is observed in most black hole X-ray binary systems.  
Like PTF11agg, the radio spectrum is flat or inverted ($f_{\nu} 
\propto \nu^{\beta}$, with $\beta \approx 0$--0.5), and 
circularly unpolarized.  However, a reasonably tight correlation 
exists between the radio and X-ray luminosity in the low-hard 
state \citep{cnf+03,gfp03}, of the form $f_{\nu}(\mathrm{radio}) \propto 
f_{\nu}(\mathrm{X-ray})^{0.7}$.  Using the derived formulation from 
\citet{gfp03} and the observed radio flux, we would expect 
an X-ray flux of $f_{X} \approx 2 \times 
10^{-12}$\,erg\,cm$^{-2}$\,s$^{-1}$ (note that this estimate is 
entirely independent of the distance to the source).  
This is more than an order of magnitude above our derived X-ray
limits.  We further note that while neutron star X-ray
binaries do not obey the same radio--X-ray correlation in the hard
state, the ratio of X-ray to radio luminosity is even larger in these
sources \citep{mbd+05}.  

Alternatively, the most luminous radio flares from LMXBs arise as the
system transitions through the intermediate state into a bright,
quasi-thermal outburst (jet emission at the highest X-ray fluxes appears to
be largely suppressed; \citealt{fbg04}).  Unlike the steady radio jets in
the low-hard phase, this state transition in the accretion flow (from
radiatively inefficient, advection-dominated to geometrically thin,
optically thick; \citealt{emn97}) can sometimes cause the ejection of
relativistic material \citep{mr99}.  While LMXBs in this state do not
always follow the same radio--X-ray correlation \citep{gfp03}, the
radio spectrum from this extended emission becomes optically thin.
The X-ray and optical fluxes can rise by several orders of magnitude on
a time scale of only a few days during these ``X-ray novae,'' but
typically both take many months to return to
quiescence \citep{ts96,cc06}.

PTF11agg differs from these X-ray novae in several major
respects.  Most importantly, to reach the intermediate (and,
ultimately, high) state where the radio flare is launched, the compact
primary must be accreting material at a substantial fraction of the
Eddington limit ($\gtrsim$ 1--10\%; \citealt{emn97}).  As a result, these outbursts
have been discovered exclusively by wide-field X-ray or $\gamma$-ray
satellites.  But for any reasonable
Galactic distance scale ($d \lesssim 10$\,kpc), our X-ray limits rule
out emission at the level of $10^{-5}$\,$L_{\mathrm{Edd}}$ (for a
1\,M$_{\odot}$ black hole or neutron star).  While our X-ray
observations were obtained $42$\,d after the initial optical
outburst, this is comparable to the $e$-folding time of these
systems.  As it requires $\sim 1$\,month for the disc mass to
accrete onto the neutron star or black hole (the viscous time
scale; \citealt{kr98}), this time delay alone cannot account for the 
many orders
of magnitude gap between our limits and the required X-ray luminosity.

In addition to the lack of bright X-ray emission, we note several
more characteristics that distinguish PTF11agg from known X-ray
nova outbursts: (1) the radio emission at late times remains
unresolved, which is difficult to reconcile with relativistic ejecta in
our Galaxy; (2) the inverted radio spectrum is inconsistent with the
optically thin emission expected at this time; (3) the time scale of
the optical decay ($\Delta t \lesssim 2$\,d) is significantly shorter
than what is observed in X-ray novae ($\tau \approx 20$--40\,d); 
and (4) the location, well off the Galactic plane ($l = 202^{\circ}$,
$b = +29^{\circ}$), is inconsistent with the known population of 
LMXBs \citep{vw95,wp96}, which have a scale height of $d_{z} 
\lesssim 1$\,kpc (although several prominent counterexamples 
are known; \citealt{thk+99,zcs+00,hmh+00,ukm+00,mdm+01,wfs+01}).

Finally, we consider systems with a white dwarf accreting material
from a (low-mass) stellar companion or another white dwarf.  
Short-timescale optical outbursts have been observed from AM CVn
systems (a white dwarf accreting H-poor material in a short-period 
orbit), but never with amplitudes larger than 4 
mag \citep{lfg+11,rbs+12}.  Dwarf novae, the analogous
phenomenon to X-ray novae in LMXBs (e.g., disc instabilities; \citealt{c93}),
have also in the past been mistaken for extragalactic
transients \citep{rsk+07}.  In fact, the amplitude and duration of the optical
outburst from PTF11agg are not unlike the most extreme dwarf novae.  
But dwarf novae rarely exhibit coincident radio emission \citep{bgm96}, 
and the few known examples have outburst durations of only 
$\sim$ 1--2 weeks \citep{bfk83,krk+08}, where the radio emission
closely follows the optical evolution.

To summarize, we have examined a variety of known classes of Galactic
transients (stellar flares, LMXBs, dwarf novae), and found that
PTF11agg does not fit neatly into any of these categories.  It should
go without saying that it is entirely possible that PTF11agg represents a
\textit{new} type of Galactic outburst, characterized by (1) bright,
rapidly fading optical emission; (2) a long-lived radio transient; and
(3) an extremely subluminous ($M_{R} \approx 11$\,mag for $d = 10$\,kpc)
quiescent optical counterpart.  Given the broad agreement between our
observations and the properties of long-duration GRB afterglows
(\S\ref{sec:fireball}), we do not further explore this possibility here.

\section{An Extragalactic Origin: Implications and Comparisons}
\label{sec:egal}
Having rejected a Galactic origin for PTF11agg, we now consider the
possibility that it resides instead at a cosmological distance (i.e.,
well beyond the Local Group and into the Hubble flow).  Assuming 
the quiescent counterpart is indeed the host galaxy of PTF11agg, we
constrain possible redshifts in \S\ref{sec:photoz}.  Even with these
crude constraints, the angular size derived in \S\ref{sec:size}
requires the presence of a relativistic outflow
(\S\ref{sec:relativistic}).  Finally, we briefly compare the observed
properties with those of known extragalactic sources capable of generating
relativistic ejecta in \S\ref{sec:egalcompare}, and quickly settle on a
long-duration GRB-like outburst (i.e., the core collapse of a massive
star) as the most plausible explanation.

\subsection{Redshift Constraints}
\label{sec:photoz}
Assuming the quiescent optical source is related to PTF11agg
(\S\ref{sec:pchance}), we can place an upper limit on its distance
based on the absence of redshifted \ion{H}{1} absorption along the line
of sight (i.e., the Lyman break).  Our $g^{\prime}$ detection implies
that redshifted Ly$\alpha$ ($\lambda_{\mathrm{rest}} = 1216$\,\AA)
falls at an observed wavelength of $\lambda_{\mathrm{Ly}\alpha}
\lesssim 4800$\,\AA\ (i.e., the middle of the $g^{\prime}$ filter
bandpass).  This results in an upper limit on the host-galaxy redshift
of $z \lesssim 3$.  

Alternatively, assuming a modest rest-frame UV luminosity for the host
galaxy ($M_{\mathrm{UV}} \lesssim -16$\,mag, or $L \gtrsim 0.01
L^{*}$; \citealt{rsp+08}), we place a lower limit on the host
redshift of $z \gtrsim 0.5$.  A similar lower limit is derived if we
compare the observed $R$-band brightness with that of known host galaxies of
long-duration GRBs \citep{jhm+12}.  

We therefore conclude that the redshift of PTF11agg should fall
somewhere in the range $0.5 \lesssim z \lesssim 3.0$.

\subsection{Evidence for Relativistic Ejecta}
\label{sec:relativistic}
In \S\ref{sec:size}, we derived two independent constraints on the
angular diameter of the emitting region from our radio observations:
$\Theta > 7$\,$\mu$as at $\Delta t_{\mathrm{obs}} \approx 42$\,d, and
$\Theta \approx 20$\,$\mu$as at $\Delta t_{\mathrm{obs}} \approx
100$\,d.  To convert these to constraints on the outflow velocity, we
use the redshift limits derived above: $0.5 \lesssim z \lesssim 3.0$
(corresponding to angular-diameter distances of 1.3--1.8\,Gpc
for a concordance $\Lambda$CDM cosmology).  Assuming ballistic
(i.e., constant velocity) expansion, the angular diameter is then
given by 
\begin{equation}
\Theta = \frac{\Gamma \beta c t}{d_{A} (1+z)},
\label{eqn:size1a}
\end{equation}
where $\Gamma$ is the outflow Lorentz factor ($\Gamma \equiv (1 -
\beta^{2})^{-1/2}$), $c$ is the speed of light, $t$ is the time since
outburst (in the observer frame), $d_{A}$ is the angular-diameter
distance, and $z$ is the source redshift.  

At $z = 0.5$, where our limits on the outflow velocity are the
weakest, we find $\Gamma > 1.2$ at $\Delta t_{\mathrm{obs}}
\approx 42$\,d, and $\Gamma \approx 1.3$ at $\Delta t_{\mathrm{obs}}
\approx 100$\,d.  These limits vary little over our redshift range of
interest, due primarily to the limited evolution of the angular-diameter
distance over this range: at $z = 3.0$, we find $\Gamma > 1.6$ at
$\Delta t_{\mathrm{obs}} \approx 42$\,d, and $\Gamma \approx 1.6$ at
$\Delta t_{\mathrm{obs}} \approx 100$\,d.

We therefore conclude that, even at this late time, the ejecta
powering the transient emission from PTF11agg are at least
transrelativistic.  For any more realistic form for the ejecta
deceleration (e.g., \citealt{bm76}), we infer that PTF11agg was
initially at least a modestly relativistic explosion.

\subsection{Comparison with Known Relativistic Sources}
\label{sec:egalcompare}
Only a handful of extragalactic sources are known to produce
relativistic ejecta: GRBs, with initial Lorentz factors as least as
large as several hundred \citep{ls01}, and possibly greater than 
$1000$ \citep{aaa+09a}; AGNs, in particular the subclass of blazars, 
with Lorentz factors as large as 50 \citep{lch+09}; and the recently 
discovered relativistic tidal disruption flares (TDFs; 
\citealt{ltc+11,bgm+11,zbs+11,bkg+11,ckh+11}), with initial Lorentz
factors $\sim 10$ \citep{mgm12,bzp+12,vkf11}.  Though with only one or
two examples to date, the known relativistic TDFs do not appear to
vary in the optical on time scales as short as those of PTF11agg, where $\delta
t \ll 1$\,d.  Furthermore, the SEDs of these sources are dominated by the
soft X-ray ($\sim 1$--10\,keV) bandpass, with peak isotropic
luminosities as large as $L_{X} \approx 10^{48}$\,erg\,s$^{-1}$.  Even
at $z = 3$, our X-ray limits (\S\ref{sec:xray}) imply $L_{X} < 6
\times 10^{45}$\,erg\,s$^{-1}$.

Blazars, however, are known to vary in the optical on short time 
scales ($\delta t < 1$\,d), and have in the past been mistaken for 
optically discovered GRB afterglows \citep{vlw+02,gof+02}.  But 
significant ($m_{p} \gtrsim 10\%$) interstellar scintillation is 
observed only very rarely in blazars ($\lesssim 1\%$ of the
population; \citealt{lrm+08}).  More importantly, the degree of
optical variability observed from PTF11agg, in particular the
amplitude from peak to quiescence ($\Delta R \gtrsim 8$\,mag
in $\Delta t \approx 1$\,month), makes this source unlikely to
belong to any known AGN class \citep{mis+11}.

On the other hand, a long-duration GRB can naturally accommodate
all of the observed properties of PTF11agg.  We find that the
standard GRB afterglow fireball model can accurately reproduce the
observed optical and radio light curves (\S\ref{sec:fireball}).  
The small initial size of the ejecta
explains the observed interstellar scintillation, though this should
be quenched as the blast wave expands relativistically (usually on a
time scale of weeks to months).  And the faint, blue  quiescent optical
counterpart is consistent with the long-duration GRB host-galaxy 
brightness distribution for $z \gtrsim 0.5$ \citep{jhm+12}.  We
therefore conclude that the most likely explanation for PTF11agg is a
long-duration GRB-like (i.e., massive star core collapse) explosion,
the first time such an event has been discovered at cosmological
distances absent a high-energy trigger.


\section{PTF11agg as a GRB: Untriggered, Orphan, or Dirty Fireball?}
\label{sec:grb}
Broadly speaking, there are three
reasons why a distant, relativistic outburst may lack detected prompt
high-energy emission.  The null hypothesis is a lack of sky
coverage (i.e., an ``untriggered'' GRB), as the more sensitive
high-energy satellites (\textit{Swift} and \textit{Fermi}) have only a
$\sim 60\%$ combined likelihood of detecting any given event
(\S\ref{sec:gamma}).  The limiting $\gamma$-ray fluence from the only
all-sky satellite available (the IPN) corresponds to an isotropic
$\gamma$-ray energy release of $E_{\gamma,\mathrm{iso}} =$ (2--200) 
$\times 10^{50}$ erg from $z = 0.5$ -- 3.0.  These values 
are not sufficiently low to rule out typical cosmological
long-duration GRBs \citep{bkb+07}, let alone the class of 
subluminous (e.g., GRB\,980425 / SN\,1998bw-like) events
uncovered in relatively nearby galaxies \citep{skn+06,cbv+06,gd07}.  
Without any additional information, the simplest explanation is that 
PTF11agg is an otherwise normal but untriggered long-duration GRB.

There exist other, more intriguing, possibilities, however.  The second
possible explanation for a GRB-like explosion absent any high-energy
signature is a viewing-angle effect.  Due to their high degree of
collimation \citep{sph99,r99}, the prompt emission from most 
GRBs is beamed away from our line of sight.  However, the 
long-lived afterglow emission may nonetheless be visible, either 
if the region generating the afterglow is less beamed than the 
$\gamma$-ray emitting material (i.e., an on-axis orphan 
afterglow; \citealt{np03}), or if, as expected, the outflow
spreads laterally at late times and illuminates an increasing fraction
of the sky (i.e., an off-axis orphan afterglow; \citealt{r97,pl98,npg02}).
The discovery of a \textit{bona fide} orphan afterglow would provide
robust constraints on the GRB beaming fraction, still a large source
of uncertainty in calculations of the true energy release and the
all-sky rate of GRBs.

Finally, a source may lack detectable high-energy emission altogether,
either because no high-energy photons were produced, or such emission
may be unable to escape to distant observers due to some internal
suppression mechanism.  It has long been noted (e.g.,
\citealt{p04}, and references therein) that the baryon composition of
the relativistic jet in the fireball model must be very finely tuned
in order to generate any detectable prompt high-energy emission (the so-called
``baryon loading problem'').  Without any baryons in the ejecta, the
internal shocks thought to power the prompt emission will not
form\footnote{An alternative possibility is that the prompt emission is
  generated by magnetic dissipation in a Poynting-flux-dominated
  outflow (see, e.g., \citealt{lb03}).}.  But
with too large a baryon fraction, the jet will not accelerate to a
sufficiently high initial Lorentz factor ($\Gamma_{0} \gtrsim 20$),
inhibiting any high-energy emission via $e^{-}$--$e^{+}$ pair
production \citep{hdl02,gng+12}.  Such explosions, dubbed 
``dirty fireballs,'' have long been predicted 
\citep{dcm00,hdl02,r03} to occur as a result of a
modest baryon loading of the jet; a proton content as small as
$M \gtrsim 10^{-4}\,{\rm M}_{\odot}$ will lower the initial Lorentz factor
sufficiently ($\Gamma_{0} \approx E_{\mathrm{KE}} / M c^{2}$),
yet can still produce the observed broad-band afterglow.
Distinguishing between a source that produces no high-energy emission
whatsoever and one in which these photons are unable to escape is
clearly challenging -- for the remainder of this work we shall refer
to such objects generically as dirty fireballs or afterglows lacking
prompt high-energy emission.

Here we attempt to discriminate between these competing hypotheses
through two different means.  First, we distinguish between
on-axis and off-axis models by comparing the observed optical and
radio emission with analytic and numerical predictions for GRB
afterglow light curves in the fireball model (\S\ref{sec:fireball}).
Second, we calculate the rate of PTF11agg-like outbursts to determine
if it is consistent with the all-sky (on-axis) GRB event rate
(\S\ref{sec:rates}).

\subsection{PTF11agg and the Fireball Model}
\label{sec:fireball}
In the standard GRB afterglow fireball model (see, e.g., \citealt{p04} 
for a review),
relativistic ejecta with (kinetic) energy $E_{\mathrm{KE}}$ sweep up 
material in the circumburst medium, forming a collisionless shock 
and accelerating electrons to a power-law distribution of energies 
with exponent $p$ and minimum Lorentz factor $\gamma_{\mathrm{m}}$.  
It is assumed that a constant fraction of the total post-shock energy 
density is partitioned to the electrons ($\epsilon_{e}$) and the
magnetic field ($\epsilon_{B}$).  These accelerated electrons then
emit synchrotron radiation, powering the long-lived X-ray, optical, 
and radio afterglow.

The observed afterglow spectrum depends on the relative ordering of
three critical frequencies: the frequency where self-absorption 
becomes important ($\nu_{\mathrm{a}}$), the characteristic
frequency of the emission ($\nu_{\mathrm{m}}$), and the frequency 
above which electrons are able to cool efficiently through radiation
($\nu_{\mathrm{c}}$). We
shall assume that all our observations occur in the ``slow'' cooling
regime ($\nu_{\mathrm{m}} < \nu_{\mathrm{c}}$), and that the
self-absorption frequency falls below the frequency range probed by
our observations ($\nu_{\mathrm{a}} < 10^{9}$\,Hz).

The light curve produced by such emission depends on the
radial profile of the circumburst medium into which the shock is expanding.
The simplest circumburst medium to consider is one in which
the density is constant ($\rho \propto r^{0}$).  This scenario is also
referred to as an interstellar medium (ISM; \citealt{spn98}), and is parametrized in
terms of the particle number density $n_{0}$, where $\rho = m_{p}
n_{0}$\,g\,cm$^{-3}$.  

Long-duration GRBs, however, have been conclusively linked to the
deaths of massive stars (e.g., \citealt{wb06}).  In the late stages of evolution,
massive Wolf-Rayet stars are stripped of their outer H and (possibly)
He envelopes in a wind, leaving behind a signature $\rho \propto 
r^{-2}$ density profile that should be discernible in the afterglow
light curve.  Wind-like environments \citep{cl00} are parametrized in 
terms of $A_{*}$, where $\rho = 5 \times 10^{11} A_{*} r^{-2}$\,g\,cm$^{-3}$.

Finally, we note that the hydrodynamical evolution also depends on the
geometry of the outflow.  GRBs are now widely believed to be
aspherical explosions \citep{r99,sph99}, biconical jets with
half-opening angle $\theta_{\mathrm{j}}$.  At early times, the jet
emission is collimated into a narrow cone ($\theta_{\mathrm{eff}} 
\approx \Gamma^{-1} \ll \theta_{\mathrm{j}}$, 
where $\Gamma$ is the Lorentz factor of the expanding shock) due to 
relativistic beaming.  As the shock slows, however, simple analytic
solutions suggest that lateral spreading of the jet becomes important, 
and on-axis observers eventually ``miss'' emission from wider angles.
This hydrodynamic transition manifests itself as an achromatic steepening
in the afterglow light curve (the ``jet break''), with an expected 
post-break decay proportional to $t^{-p}$.  While more recent
numerical simulations have suggested a more complex picture
of the jet-break phenomenon \citep{zm09,vzm10,gp12}, the assumption 
of a general light-curve steepening around the time when 
$\Gamma = 1 / \theta_{\mathrm{j}}$ remains largely valid.

At $\Delta t \approx 0.2$\,d (the approximate time of discovery), we
expect the optical bandpass to fall below the cooling frequency (i.e.,
$\nu_{\mathrm{opt}} < \nu_{\mathrm{c}}$).  If we also assume that
these early optical data occur before any jet break (see below), the
observed temporal decay index ($\alpha_{\mathrm{opt}} = 1.66 \pm
0.35$) can be translated directly into the electron spectral index
$p$.  For a constant-density medium, we find $p = 3.21 \pm 0.47$,
while for a wind-like environment, we infer $p = 2.55 \pm 0.46$.
Electron spectral indices derived from previous observations
of long-duration GRBs \citep{skr06,svr+08,ced+10} fall in the range
$\sim 2$--3, so the large
uncertainty makes it difficult to distinguish between the competing
density profiles solely on this basis.

While the radio emission is relatively variable at $\Delta t \gtrsim
40$\,d, the approximate radio spectral index, $\beta_{\mathrm{radio}}
\approx 0.3$, implies that the peak synchrotron frequency
$\nu_{\mathrm{m}}$ is not well below the radio at this time (or else we
would expect $\beta_{\mathrm{radio}} \approx -1$).  Conservatively, we
assume $\nu_{\mathrm{m}} (\Delta t = 40\,\mathrm{d}) \gtrsim 10$\,GHz,
and $f_{\nu_{\mathrm{m}}} (\Delta t = 40\,\mathrm{d}) \gtrsim
300$\,$\mu$Jy.  For a wind-like medium, $\nu_{\mathrm{m}} \propto
t^{-3/2}$ and $f_{\nu_{\mathrm{m}}} \propto t^{-1/2}$.
Extrapolating back to the time of optical discovery, we conclude $\nu_{m}
(\Delta t = 0.2\,\mathrm{d}) \gtrsim 3 \times 10^{13}$\,Hz, and
$f_{\nu_{\mathrm{m}}} (\Delta t = 0.2\,\mathrm{d}) \gtrsim 4 \times
10^{3}$\,$\mu$Jy.  For $\nu_{\mathrm{m}} < \nu < \nu_{\mathrm{c}}$,
$f_{\nu} \propto \nu^{(1-p)/2} \approx \nu^{-0.77}$.  Thus, we find that 
the inferred optical ($R$-band) flux at discovery, $f_{\nu} \gtrsim 
500$\,$\mu$Jy, is a factor of $\sim 3$ larger than our 
observations at this time.  For a constant-density environment, 
the peak flux is constant in time, and so a similar analysis yields 
a self-consistent result.  We therefore do not consider a wind-like 
medium any further.

\begin{figure}[t!]
  \plotone{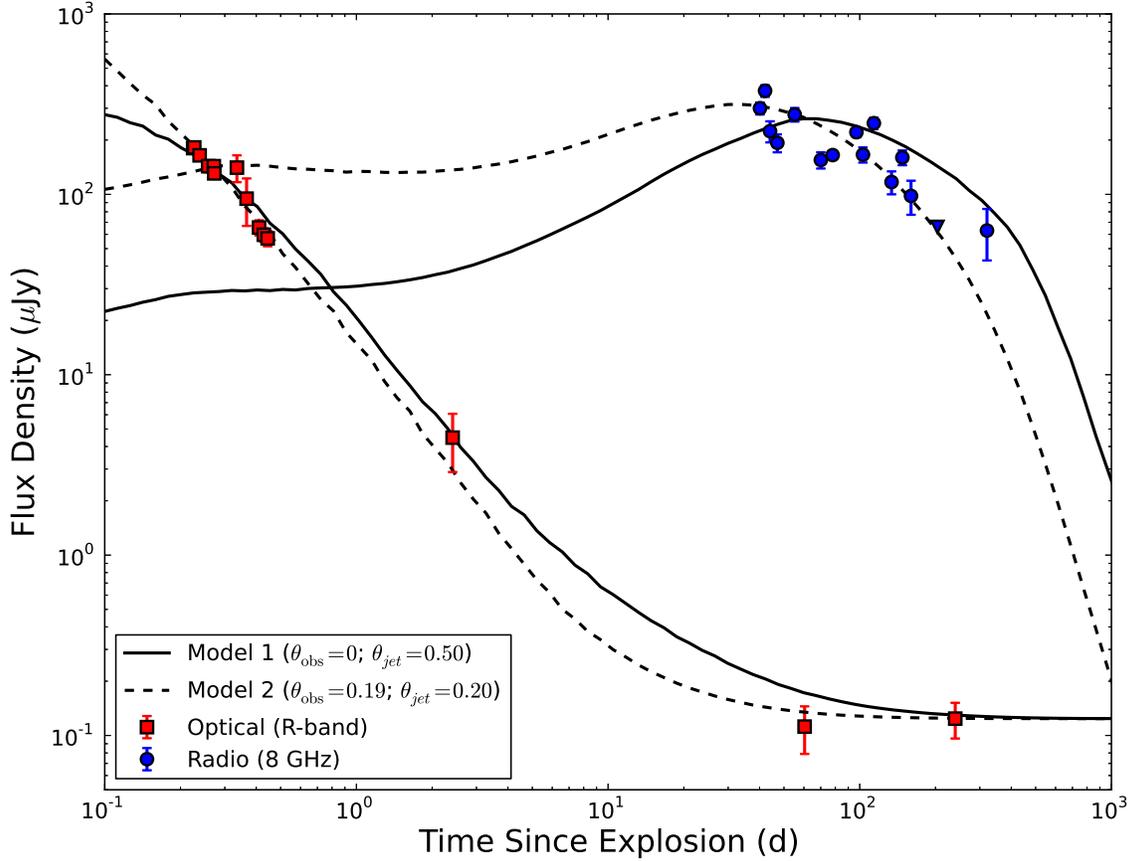}
  \caption[]{\footnotesize
  The solid line shows the best-fit afterglow 
  model \citep{vvm12} when
  the relativistic jet is oriented directly along the line of sight to
  the observer (i.e., $\theta_{\mathrm{obs}} = 0$).  The dashed curve
  displays the best-fit model when the viewing angle is allowed to
  vary freely.  Given the relatively sparse dataset (in particular
  the lack of X-ray observations), a wide variety of models are able
  to reproduce the observed optical and radio emission.  However, we
  find it impossible to reproduce the observed emission when the
  viewing angle is outside the cone of the jet (i.e.,
  $\theta_{\mathrm{obs}} > \theta_{\mathrm{j}}$).  We therefore
  consider it unlikely that viewing angle alone can account for the
  lack of high-energy emission from PTF11agg.
  }
\label{fig:agmodel}
\end{figure}

Using these general constraints, we have used the software described
by \citet{vvm12} to fit the observed optical
and radio light
curves to afterglow models calculated from high-resolution 
two-dimensional relativistic hydrodynamical jet simulations.  In
all cases, we have assumed a constant-density circumstellar medium 
and adopted a fiducial redshift of 1.  We find that a relatively wide 
set of parameters is able to reproduce the 
observations\footnote{In all cases the predicted X-ray flux is well
  below the XRT limit (\S\ref{sec:xray}), so we have not included this
  point in our fitting.}, largely consistent with values derived from previous 
GRB afterglow modeling \citep{pk01a,pk01b,yhs+03}, although with a 
somewhat smaller circumburst density ($n_{0} \lesssim 0.1$ cm$^{-3}$).  
The best-fit model, assuming the observer is oriented directly along the jet
axis (i.e., $\theta_{\mathrm{obs}} = 0$), is plotted in Figure~\ref{fig:agmodel}
($E_{\mathrm{KE}} = 3 \times 10^{52}$\,erg, $\theta_{\mathrm{j}} =
0.50$\,rad, $n_{0} = 1 \times 10^{-3}$\,cm$^{-3}$, $\epsilon_{e} =
0.1$, $\epsilon_{B} = 0.1$, and $p = 2.9$).  Repeating a similar analysis
with the software described by \citet{yhs+03} yields
qualitatively similar results.  The general agreement between the
fireball model and our optical and radio observations 
supports our conclusion that PTF11agg is most likely a distant,
relativistic explosion like other long-duration GRBs.

One concern regarding the $z = 1$ models is the implied angular size
of the source ($\Theta$).  In the best-fit on-axis model, we find that 
the outflow has an angular size $\Theta \approx 40$\,mas at $\Delta t =
40$\,d.  This is somewhat larger than the value inferred from our
scintillation analysis, but within a factor of two.  However, many of
the other $z = 1$ models that provide a reasonable fit to the data   
are likely to be too spatially extended to scintillate strongly
at $\Delta t \approx 100$\,d.  One potential solution may be a more
distant origin; due to cosmological time dilation, an observer-frame
time of $\Delta t = 100$\,d would correspond to only 25 rest-frame days
post-explosion at $z = 3$, half the expansion time as inferred at $z
= 1$.  Given the large spread in acceptable models, however, we do not
explore this possibility further here.

Next we consider limits on the opening angle and observer
orientation from the observed optical and radio emission.  
After the jet 
break, the peak synchrotron flux declines linearly with time (in a 
constant-density environment).  Thus, if the jet break occurred 
well before the first radio observations, the large radio flux would 
be difficult to reconcile with our early optical observations.  We
consider it likely, then, that $t_{\mathrm{j}} \gtrsim 40$\,d.  
These conclusions are largely confirmed by our numerical models, where
we find that the opening angle is only weakly constrained to be 
$\theta_{\mathrm{j}} \gtrsim 0.15$\,rad.

In addition to cases where the observer is oriented directly along the
jet axis (i.e., $\theta_{\mathrm{obs}} = 0$), we also have considered
more general geometries, where the observer may be oriented off-axis,
either within (i.e., $\theta_{\mathrm{obs}} < \theta_{\mathrm{j}}$) or
outside ($\theta_{\mathrm{obs}} > \theta_{\mathrm{j}}$) the jet
opening angle.  For simplicity, 
we consider only ``top-hat'' jet geometries, where the
jet Lorentz factor is given by a step function.
For $\theta_{\mathrm{obs}} > \theta_{\mathrm{j}}$,
observers will see rising emission until approximately the time of the
jet break, after which the decay will resemble the on-axis case.  For
observers off-axis but within the jet opening angle, the modifications
to the on-axis afterglow light curves will be more 
subtle \citep{gpk02,zm09,vzm10}. 

The arguments used above to infer $t_{\mathrm{j}} \gtrsim 40$\,d
necessarily require that the observer cannot be well outside the jet
opening angle (or else we would expect to see post jet-break decay).
This result is borne out by our numerical modeling, where geometries
with $\theta_{\mathrm{obs}} > \theta_{\mathrm{j}}$ are unable to
accurately reproduce the observed light curves.  Allowing the observer
orientation to vary as a free parameter, the best-fit afterglow model
is plotted as a dashed line in Figure~\ref{fig:agmodel} 
($E_{\mathrm{KE}} = 9 \times 
10^{52}$\,erg, $\theta_{\mathrm{j}} = 0.20$\,rad, $\theta_{\mathrm{obs}} =
0.19$\,rad, $n_{0} = 1 \times 10^{-3}$\,cm$^{-3}$, $\epsilon_{e} =
0.04$, $\epsilon_{B} = 0.2$, and $p = 3.0$).

Finally, we can derive a lower limit on the distance to PTF11agg based
only on the observed radio evolution.  The radio spectrum at $\Delta t
= 67$\,d, $f_{\nu} \propto \nu^{\beta}$, with $\beta \approx 0.3$, is 
inconsistent with Sedov-Taylor blast-wave evolution.  
In other words, the outgoing shock wave has not transitioned to 
nonrelativistic expansion at this point in time.  The nonrelativistic
transition will occur at a time \citep{wwf11}
\begin{equation}
t_{\mathrm{nr}} = 1100 \left(
  \frac{E_{\mathrm{KE}}}{10^{53}\,\mathrm{erg}} \right) \left(
  \frac{n_{0}}{1\,\mathrm{cm}^{-3}} \right)\,\mathrm{d}.
\label{eqn:tnr}
\end{equation}
Even neglecting our previous finding of a low circumburst density, we
infer a sizeable lower limit on the blast-wave kinetic energy:
$E_{\mathrm{KE}} \gtrsim 10^{50}$\,erg.  Integrating over the observed
8\,GHz radio light curve, we measure a fluence of $S_{\mathrm{rad}} =
2 \times 10^{-10}$\,erg\,cm$^{-2}$.  For a typical cosmological
distance ($z = 1$, or $d_{L} = 2 \times 10^{28}$\,cm), this
corresponds to a radiated energy of $E_{\mathrm{rad}} = 3 \times
10^{47}$\,erg, a typical radiative efficiency for a GRB.  But for a
Galactic outburst, the radiated energy would be many orders of
magnitude smaller ($E_{\mathrm{rad}} = 6 \times 10^{35}$\,erg at $d =
10$\,kpc).  Unless the radiative efficiency was incredibly small, we
once again conclude that PTF11agg must lie at a cosmological distance.

\subsection{The Rate of PTF11agg-like Events}
\label{sec:rates}
Our objective in this section is to estimate the number of GRB 
optical afterglows discovered by chance (i.e., not as a result of 
deliberate follow-up observations of a high-energy trigger) by 
PTF.  If the likelihood of chance detection of an untriggered
afterglow with PTF is significant, we will conclude that the rate of
PTF11agg-like events is consistent with the rate of normal (i.e.,
on-axis) long-duration GRBs.  If this probability is small, then
we can use these calculations to place lower limits on the observed
frequency of PTF11agg-like outbursts (in units of the GRB rate). 
Given the relatively complex nature of PTF scheduling \citep{lkd+09}, 
we have conducted a series of Monte Carlo simulations to this end.

PTF began full operations on about 2009 April 1.  We have
retrieved a listing of all images obtained beginning at this time
through 2012 December 31, or over a period of 45 months.  
We removed fields at
Galactic latitude $|b| < 20^{\circ}$ (due to the large foreground
extinction)\footnote{Given that the primary objective of PTF is the
  discovery of extragalactic transients, this represents less than
  10\% of the total number of observations.}.  
The resulting sample includes 129,206 pointings, each
covering an area of 7.2\,deg$^{2}$.  The sample comprises 1940 unique
fields, each imaged an average of 67 times.  

Since launch, the \textit{Swift} BAT\footnote{See
  http://swift.gsfc.nasa.gov/docs/swift/archive/grb\_table .} 
detects GRBs at a rate of $\approx 90$\,yr$^{-1}$.
The field-of-view of the BAT
is $\sim 2$\,sr, and the instrument has a duty cycle of $\sim
90\%$.  Thus, the all-sky rate for events at the BAT threshold is 
$\sim 630$\,yr$^{-1}$.  Over the 3.75 yr period of interest, 
the total number of all-sky GRBs is $\sim 2360$.  
We note that this is an upper limit to the long-duration 
GRB rate, as we have included short-duration GRBs in this sample as well.

For each trial, we create a mock catalog of 2360 GRBs.  Each GRB is 
randomly assigned a trigger time $t_{0}$ (uniformly distributed 
between 2009 April 1 and
2012 December 31) and spatial coordinates $\alpha$, $\delta$
(isotropically distributed on the sky).  To estimate the duration over 
which the optical afterglow would be detectable by PTF, we utilize the
sample of 29 long-duration afterglows from the Palomar 60\,inch (P60) 
\textit{Swift} afterglow catalog \citep{ckh+09}.  These events were
selected solely on the basis of visibility to Palomar Observatory, 
so they should represent an unbiased sample of
the \textit{Swift} afterglow brightness distribution.  For each
event in the P60-\textit{Swift} sample, we have calculated the
amount of time following the high-energy trigger that the afterglow is 
brighter than $R = 20$\,mag.  These values
range from $< 204$\,s (GRB\,050721) to 1.2\,d (GRB\,050820A).  Each
mock GRB is randomly assigned one of the 29 actual ``visibility 
windows'' from this sample\footnote{For GRBs without any detected
  optical afterglow (e.g., ``dark'' bursts), we use the earliest
  non-detection below our sensitivity threshold for the visibility window.
  If anything, this would bias us to over-estimate the expected number
  of untriggered GRB afterglow detections by PTF.}.

For each mock GRB, we then determine if the event occurred within the
7.2\,deg$^{2}$ footprint of any individual PTF image, and, if so, if
the time of observation occurred within the necessary window during
which the afterglow was brighter than 20\,mag.  The number
of afterglows detected in each trial ($N_{\mathrm{GRB}}$), together
with the number of individual frames on which each detected afterglow was 
brighter than the P48 sensitivity limit ($N_{\mathrm{Det}}$), were
then recorded.  The results of 1000 individual runs (i.e., different 
randomly selected groups of 2360 GRBs) constitute a sufficiently 
large sample to evaluate the likelihood of serendipitous detection 
of long-duration GRB afterglows with PTF.

\begin{figure}[t!]
  \plotone{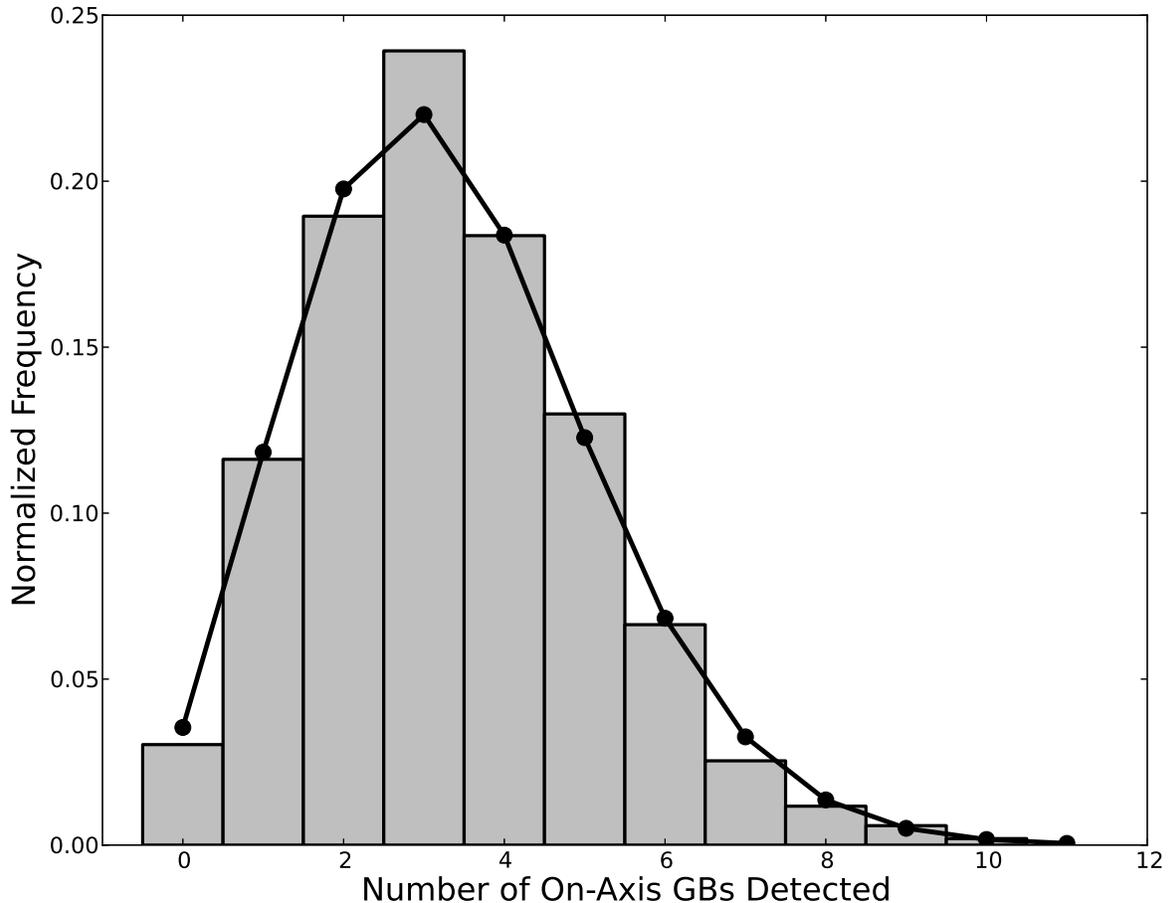} 
  \caption{\footnotesize
    Normalized histogram of the number of serendipitous
    detections of normal on-axis GRB afterglows by 
    PTF in our 1000 Monte Carlo trials.  The distribution is reasonably well
    described by a Poisson function with $\lambda = 3.3$ (solid black line).}
\label{fig:NGRB}
\end{figure}

In the 1000 trials conducted, at least one GRB afterglow was detected
(i.e., $N_{\mathrm{GRB}} \ge 1$) in 970 instances. 
Thus, the probability of detecting at least one 
on-axis afterglow over the course of the first two years of PTF is 
quite high, $P(N_{\mathrm{GRB}} \ge 1) = 97\%$.  The
expectation value for the number of afterglows detected is $\lambda = 
3.3$.  The distribution
of the number of afterglows detected in our 1000 trials
is reasonably well described 
by Poisson statistics (Figure~\ref{fig:NGRB}).  In this respect, 
then, PTF11agg appears to be consistent with a normal on-axis GRB.

However, the field in which PTF11agg was identified (the Beehive
cluster) is atypical amongst PTF pointings.  Most fields are
only observed 2 or 3 times per night (multiple images are used to identify
Solar-System objects).  But the Beehive cluster is a ``high-cadence''
field, observed many times ($\gtrsim 10$) per night during its
observing season.  Instead of calculating the rate of 
afterglow detections over the entire survey (i.e.,
$N_{\mathrm{GRB}}$), a more appropriate
comparison would limit the scope to similar high-cadence fields.

We therefore consider on how many individual 
images each of the $3340$ ``detected'' GRB afterglows (in our 1000 trials)
were above the P48 limiting magnitude (i.e., $N_{\mathrm{Det}}$).  This is 
illustrated in Table~\ref{tab:Ndet}.  The vast majority of the
afterglows are detected on only one or two images (87\%).  In fact, 
in our 1000 trials, an optical afterglow was detected on at least ten 
individual images only 11 times (i.e., $P(N_{\mathrm{Det}} \ge 10) =
2.6\%$).  PTF11agg was detected 11 times on 2011 January 30 with 
$R < 20$\,mag.

We can understand this result analytically in the following manner.
In the case where the integration time ($\delta t$) is much smaller
than the period over which a transient is visible ($\tau$),
the number of detectable events at any given time will be
\begin{equation}
q = \frac{ \Omega \mathcal{N} \tau }{4 \pi ,}
\end{equation}
where $\Omega$ is the field of view (in steradian) and 
$\mathcal{N}$ is the all-sky
event rate.  For long-duration GRB optical afterglows, serendipitous
detection by PTF will be dominated by the $\sim 10\%$ of events that remain
brighter than $R < 20$\,mag for $\tau \approx 1$\,d (certainly this is
true for those afterglows with $N_{\mathrm{Det}} > 3$).  Thus, for the
PTF project, $\Omega / 4 \pi = 1.7 \times 10^{-4}$\,sr (7.2\,deg$^{2}$), and
adopting $\mathcal{N} \approx$ 0.1 $\times$ 630\,yr$^{-1}$ and $\tau
\approx 2.7 \times 10^{-3}$ (1\,d), we find $q \approx 3.0 \times
10^{-5}$ events per field.

The expected number of detected events, $\lambda$, will then be $q 
N_{\mathrm{Obs}}$, where $N_{\mathrm{Obs}}$ is the number of
(independent) measurement epochs.  Over the two-year period of
interest, the number of individual P48 images obtained is
$N_{\mathrm{Obs}}($all$) = 1.3 \times 10^{5}$.  Thus, we predict 
$\lambda \approx 3.7$, in good agreement with the results of our 
Monte Carlo simulations.

Conversely, we can calculate the relative frequency of high-cadence
($N_{\mathrm{Obs}}[> 10]$) observations in our two-year PTF 
sample by measuring how often each field was observed on a 
nightly basis.  The results of this analysis are shown in the 
far-right column of Table~\ref{tab:Ndet}.  As is evident,
high-cadence observations with $N_{\mathrm{Obs}}$($> 10$) (i.e., more 
than 10 observations of a field obtained in a single night) occur with a
frequency of 1\% when compared with regular-cadence fields
($N_{\mathrm{Obs}}[1] + N_{\mathrm{Obs}}[2]$).  

From this analysis, we conclude that the rate of PTF11agg-like events
is inconsistent with the rate of long-duration GRBs with 97.4\%
confidence.  Admittedly a number of assumptions went into this
analysis, and one should always be careful with results drawn from
such an \textit{a posteriori} analysis.  But independent of the exact
likelihood, we conclude that the probability of untriggered afterglow
detection in a high-cadence PTF field is small.
Either we have been quite lucky, or we may have uncovered a new, more
common class of distant, relativistic outbursts lacking entirely in
high-energy emission.

It is crucial to verify, however, that our inferred rate does not
violate any other limits on short-timescale transients, either from
PTF itself, or from previous optical and radio surveys.  As
highlighted above, low-cadence fields are observed significantly more
frequently with PTF than high-cadence fields like the Beehive.  Thus,
any short-timescale ($\Delta t \lesssim 1$\,d) transient should be
detected in many more $N_{\mathrm{Obs}}(1)$ and $N_{\mathrm{Obs}}(2)$ 
fields than high-cadence fields.  In the case of PTF11agg,
repeating the above Monte Carlo simulations for a transient population
with five times the GRB event rate (but the same optical brightness
distribution), we find an expected number of detected sources of
$\lambda = 16.7$ in all fields.  At first glance, the fact that we have not
discovered such a population of sources would seem to favor the
untriggered GRB scenario.  

Here it is important to distinguish between transient detection, by
which we mean a source is above the P48 sensitivity limit on a given
image, and discovery, where a transient is flagged as astrophysically
interesting (by software or human beings; \citealt{brn+11}).  Because of
the large number of uncatalogued asteroids near the PTF limit, our
software requires at least two detections at a given
location to flag a source as a {\it bona fide} transient (e.g., to
``discover'' the source).  Thus, any PTF11agg-like outburst with only
a single detection ($N_{\mathrm{Det}} = 1$) will never be discovered
by our survey.  Likewise, there may be subtle biases limiting our
capability to identify and/or conduct follow-up observations of 
similar short-timescale transients with only a few detections.

Whether these discovery biases are sufficient to account for the lack
of similar sources in our low-cadence fields with PTF remains to be
seen.  We have attempted to search through all PTF discoveries that
were detected only on a single night (independent of
$N_{\mathrm{Det}}$)\footnote{We cannot avoid the requirement of at
  least two detections, however.  Otherwise we would be completely
  swamped with asteroids, which are detected at a rate of thousands
  per night.}, but have yet to uncover any additional viable
candidates.  Ultimately, future wide-field, high-cadence optical
surveys may be required to resolve this issue.

Finally, we compare our derived rate of PTF11agg-like events with
previous searches for orphan 
optical \citep{vlw+02,bwb+04,raa+05,rgs06,rok+08} and 
radio \citep{low+02,bkf+03,gop+06,snb+06} afterglows, to verify that our
results are consistent with these limits.  The tightest constraints on
the rate of relativistic outbursts come from radio surveys, where
\citet{gop+06} derive a limit on the all-sky volumetric rate of
GRB-like explosions of $\dot{N} < 
10^{3}$ events Gpc$^{-3}$ yr$^{-1}$.  Even assuming an all-sky GRB rate as
large \citep{gd07} as 100\,Gpc$^{-3}$\,yr$^{-1}$ (more recent
estimates suggest a significantly smaller value; \citealt{bbp10}), a
population of PTF11agg-like events occurring at a rate of $\sim 5$
times that of normal GRBs is consistent with these results.  Our
derived rate is therefore orders of magnitude lower than the all-sky
rate of Type Ibc supernovae ($\dot{N} = 2.6 \times
10^{4}$\,Gpc$^{-3}$\,yr$^{-1}$; \citealt{lcl+11}).  It may approach the rate of
low-luminosity GRBs \citep{skn+06,cbv+06,gd07}, although this depends
both on the assumed beaming correction and the true GRB rate.

\section{Summary and Conclusions}
\label{sec:disc}
To summarize our results, we report here the discovery of 
PTF11agg, a rapidly
fading optical transient with a long-lived, scintillating radio
counterpart.  Together with the observed optical and radio light
curves, the detection of a faint, blue quiescent counterpart at the
location of PTF11agg indicates that the transient likely originated in
the distant universe.  Using our measurements of the source size
derived from the radio observations, we infer that PTF11agg must be
powered by a relativistic outflow.  These properties are all
consistent with the population of long-duration GRB afterglows,
marking the first time such an event has been discovered at
cosmological distances without a high-energy trigger.

Searching various high-energy satellites, we find no potential
$\gamma$-ray counterpart for PTF11agg.  We therefore consider three
possible explanations that can simultaneously account for a GRB-like
explosion without any associated prompt high-energy emission: an
untriggered GRB, an orphan afterglow, and a dirty fireball.

Using the all-sky rate of GRBs discovered by the \textit{Swift}
satellite, together with a measurement of their observed optical
brightness distribution, we have calculated the
likelihood of serendipitous untriggered GRB afterglow detection by PTF
(April 2009 --
December 2012).  Surprisingly, we found that the \textit{a posteriori}
probability of untriggered GRB afterglow detection in a high-cadence
field like the one where PTF11agg was found (11 observations on a
single night) is only 2.6\%.  While we cannot rule out entirely our
null hypothesis that PTF11agg is an untriggered GRB, this probability 
is sufficiently low that we consider alternative interpretations as well.

The afterglow emission from an orphan GRB will rise in flux
at early times, as more and more of the jet becomes visible due to relativistic
beaming effects.  Using both analytic and numerical
formulations, we are unable to reproduce the observed PTF11agg light
curves unless the observer viewing angle is within the opening angle
of the jet.  While these models assume a relatively simple jet
structure, the requirement of rising afterglow emission at early times 
is a robust prediction for all off-axis models.

A more intriguing possibility is that PTF11agg may represent a new
class of relativistic outbursts with little or no corresponding
high-energy emission.  In much the same way that SN\,2009bb
\citep{scp+10} demonstrated that the more nearby, subluminous class 
of GRBs may generate relativistic ejecta yet lack high-energy
emission, PTF11agg may play an analogous role for the more energetic, 
cosmologically distant sample of long-duration GRBs.  Dirty fireballs
(i.e., a baryon-loaded jet) are one possible explanation, though
alternative possibilities surely exist as well.

In this picture, the inferred rate of PTF11agg-like events must be four times higher
(90\% confidence) than the rate of on-axis long-duration GRBs.  
When combined with traditional
core-collapse supernovae and long-duration GRBs, these objects would
enable a more complete census of the deaths of massive stars,
and also provide a probe of the location of massive-star
formation in distant galaxies without the need for a high-energy
satellite trigger.

Regardless of its ultimate origin, we expect such sources to be
discovered in large numbers by ongoing and future wide-field,
high-cadence optical surveys such as the Catalina Real-Time Transient
Survey \citep{ddm+09}, PTF, Pan-STARRS \citep{kbc+10}, 
and the Large Synoptic Survey Telescope \citep{ita+08}.  Furthermore,
the discovery of PTF11agg bodes well for optical surveys in the future
era of gravitational wave astronomy, as the electromagnetic
counterparts of gravitational wave sources should exhibit largely
similar observational signatures (though they are also expected to be
associated with more nearby galaxies; \citealt{np11,mb12}).

\acknowledgments
We wish to thank David Levitan and
Kunal Mooley for obtaining observations used in this work, and John
Tomsick for valuable comments on the manuscript.  We are grateful to the
following IPN team members for sharing their data:
K.~Hurley, I.~G.~Mitrofanov, D.~Golovin, M.~L.~Litvak, A.~B.~Sanin,
W.~Boynton, C.~Fellows, K.~Harshman, R.~Starr,
S.~Golenetskii, R.~Aptekar, E.~Mazets, V.~Pal'shin, D.~Frederiks, D.~Svinkin,
A.~von Kienlin, X.~Zhang, K.~Yamaoka, T.~Takahashi,
M.~Ohno, Y.~Hanabata, Y.~Fukazawa, M.~Tashiro, Y.~Terada, T.~Murakami,
K.~Makishima, T.~Cline , S.~Barthelmy, J.~Cummings, N.~Gehrels, H.~Krimm,
D.~Palmer, J.~Goldsten, V.~Connaughton, M.~S.~Briggs, and
C.~Meegan.

A.V.F.~and his group acknowledge generous financial assistance
from Gary \& Cynthia Bengier, the Richard \& Rhoda Goldman Fund,
the Christopher R. Redlich Fund,
NASA/{\it Swift} grants NNX10AI21G and NNX12AD73G, the
TABASGO Foundation, and NSF grant
AST-1211916.  A.C. acknowledges support from LIGO, which was
constructed by the California Institute of Technology and the
Massachusetts Institute of Technology with funding from the NSF
and operates under cooperative agreement PHY-0757058.
D.A.P.~is supported by NASA through Hubble Fellowship
grant HST-HF-51296.01-A awarded by the Space Telescope Science
Institute, which is operated by the Association of Universities for
Research in Astronomy, Inc., for NASA, under contract NAS 5-26555.
Research by A.G.Y.~and his team is supported by grants from the 
ISF, BSF, GIF, the EU/FP7 via an ERC grant, and a Kimmel Award. 
P.J.G. acknowledges support
from Caltech during his 2011 sabbatical stay.  E.O.O. is incumbent of
the Arye Dissentshik career development chair and
is grateful to support by a grant from the Israeli Ministry of Science.
A.A.M.~is supported by the NSF
Graduate Research Fellowship Program. J.S.B.~acknowledges 
NSF grant CDI-0941742.
M.M.K~acknowledges generous support from the Hubble Fellowship and
Carnegie-Princeton Fellowship.
D.P. is grateful for the AXA research fund.
A.S.~is supported by a Minerva Fellowship.  

Observations were obtained with the Samuel Oschin telescope and the
Hale telescope at Palomar Observatory as part of the Palomar Transient Factory
project, a scientiﬁc collaboration between the California Institute of
Technology, Columbia University, Las Cumbres Observatory, the
Lawrence Berkeley National Laboratory, the National Energy Research
Scientiﬁc Computing Center, the University of Oxford, and the
Weizmann Institute of Science.  The National Energy Research
Scientific Computing Center, supported by the Office of Science of
the U.S. Department of Energy, provided staff,
computational resources, and data storage for this project.
Support for CARMA construction was derived from the Gordon
and Betty Moore Foundation, the Kenneth T.~and Eileen L.~Norris
Foundation, the James S.~McDonnell Foundation, the Associates
of the California Institute of Technology, the University of Chicago,
the states of California, Illinois, and Maryland, and the NSF.
Ongoing CARMA development and operations are
supported by the NSF under a cooperative
agreement, and by the CARMA partner universities.
Some of the data presented herein were obtained at the W. M. Keck
Observatory, which is operated as a scientific partnership among the
California Institute of Technology, the University of California and
NASA; the Observatory
was made possible by the generous financial support of the W. M. Keck
Foundation.  PAIRITEL is operated by the Smithsonian Astrophysical
Observatory (SAO) and was made possible by a grant from
the Harvard University Milton Fund, a camera loan from the
University of Virginia, and continued support of the SAO and
UC Berkeley. The PAIRITEL project is further supported by
NASA/{\it Swift} Guest Investigator grant NNX08AN84G.
This work made use of data supplied by the UK {\it Swift} Science
Data Centre at the University of Leicester. It also
made use of the NASA/IPAC Extragalactic Database (NED),
which is operated by the Jet Propulsion Laboratory, California
Institute of Technology, under contract with NASA.
In addition, we have utilized the SIMBAD
database, operated at CDS, Strasbourg, France.

{\it Facilities:} \facility{PO: 1.2m (PTF)}, \facility{Hale (WIRC)},  \facility{Keck:I (LRIS)},
\facility{Magellan: Baade (IMACS)}, \facility{FLWO: 2MASS (PAIRITEL)},
\facility{VLA}, \facility{CARMA},
\facility{Fermi (GBM)}, \facility{Swift (BAT, XRT)}



\clearpage
\begin{deluxetable}{lcccc}
\tabletypesize{\scriptsize}
\tablecaption{Optical/Near-Infrared Observations of PTF11agg}
\tablewidth{0pt}
\tablehead{
\colhead{Date} & \colhead{Telescope/Instrument} & \colhead{Filter} &
\colhead{Exposure Time} & \colhead{Magnitude} \\
\colhead{(MJD)} & & & (s) & 
}
\startdata
55590.30519 & P48/CFHT12k & $R$ & 540 & $> 21.9$ \\
55591.22026 & P48/CFHT12k & $R$ & 60 & $18.26 \pm 0.05$ \\
55591.22245 & P48/CFHT12k & $R$ & 60 & $18.25 \pm 0.04$ \\
55591.23391 & P48/CFHT12k & $R$ & 60 & $18.36 \pm 0.05$ \\
55591.25326 & P48/CFHT12k & $R$ & 60 & $18.51 \pm 0.08$ \\
55591.26691 & P48/CFHT12k & $R$ & 60 & $18.51 \pm 0.04$ \\
55591.26800 & P48/CFHT12k & $R$ & 60 & $18.61 \pm 0.06$ \\
55591.33081 & P48/CFHT12k & $R$ & 60 & $18.53 \pm 0.17$ \\
55591.36188 & P48/CFHT12k & $R$ & 60 & $18.96 \pm 0.28$ \\
55591.40604 & P48/CFHT12k & $R$ & 60 & $19.36 \pm 0.10$ \\
55591.42439 & P48/CFHT12k & $R$ & 60 & $19.46 \pm 0.09$ \\
55591.43978 & P48/CFHT12k & $R$ & 60 & $19.51 \pm 0.10$ \\
55593.40775 & P48/CFHT12k & $R$ & 420 & $22.15 \pm 0.33$ \\
55594.23819 & P48/CFHT12k & $R$ & 300 & $> 21.2$ \\
55621.19100 & PAIRITEL & $H$ & 2246 & $> 20.4$ \\
55621.19100 & PAIRITEL & $J$ & 2246 & $> 20.6$ \\
55621.19100 & PAIRITEL & $K_{s}$ & 2246 & $> 19.7$ \\
55624.49 -- 55678.28 & Keck I/LRIS & $g^{\prime}$ & 6680 & $26.63 \pm 0.33$ \\
55624.49 -- 55678.28 & Keck I/LRIS & $R$ & 5700 & $26.28 \pm 0.28$ \\
55830.60259 & Keck I/LRIS & $g^{\prime}$ & 2100 & $26.34 \pm 0.19$ \\
55830.59849 & Keck I/LRIS & $R$ & 2160 & $26.17 \pm 0.22$ \\
55944.22461 & Magellan/IMACS & $I$ & 2400 & $> 25.2$ \\
56014.27324 & P200/WIRC & $K_{s}$ & 1200 & $> 22.6$ \\
\enddata
\label{tab:opt}
\end{deluxetable}


\clearpage
\begin{deluxetable}{lcccc}
\tabletypesize{\scriptsize}
\tablecaption{Radio Observations of PTF11agg}
\tablewidth{0pt}
\tablehead{
\colhead{Date} & \colhead{Observatory} & \colhead{Frequency} &
\colhead{Integration Time} & \colhead{Flux Density} \\
\colhead{(2011 UT)} & & \colhead{(GHz)} & \colhead{(min)} &
\colhead{($\mu$Jy)}
}
\startdata
Mar. 11.27 & VLA & 8.46 & 15.5 & $300 \pm 23$ \\
Mar. 13.12 & VLA & 4.50 & 15.5 & $217 \pm 34$ \\
Mar. 13.12 & VLA & 7.92 & 15.5 & $375 \pm 28$ \\
Mar. 14.05 & CARMA & 93.5 & 492.0 & $450 \pm 140$ \\
Mar. 15.08 & VLA & 4.50 & 15.5 & $183 \pm 36$ \\
Mar. 15.08 & VLA & 7.92 & 15.5 & $224 \pm 30$ \\
Mar. 18.08 & VLA & 4.50 & 15.5 & $58 \pm 37$ \\
Mar. 18.08 & VLA & 7.92 & 15.5 & $171 \pm 31$ \\
Mar. 18.10 & VLA & 22.46 & 10.8 & $237 \pm 79$ \\
Mar. 18.15 & VLA & 4.50 & 15.5 & $93 \pm 35$ \\
Mar. 18.15 & VLA & 7.92 & 15.5 & $215 \pm 29$ \\
Mar. 18.17 & VLA & 22.46 & 10.8 & $460 \pm 77$ \\
Mar. 26.22 & VLA & 8.46 & 15.5 & $277 \pm 24$ \\
Apr. 5.17 & VLA & 21.8 & 30.1 & $271 \pm 18$ \\
Apr. 7.03 & CARMA & 93.5 & 378.0 & $480 \pm 120$ \\
Apr. 10.11 & VLA & 4.8 & 13.7 & $127 \pm 21$ \\
Apr. 10.11 & VLA & 7.4 & 13.7 & $155 \pm 16$ \\
Apr. 11.01 & CARMA & 93.5 & 342.0 & $-40 \pm 150$ \\
Apr. 18.15 & VLA & 4.8 & 37.2 & $81 \pm 11$ \\
Apr. 18.15 & VLA & 7.4 & 37.2 & $165 \pm 8$ \\
May 6.08 & VLA & 4.8 & 13.7 & $232 \pm 18$ \\
May 6.08 & VLA & 7.4 & 13.7 & $221 \pm 14$ \\
May 13.15 & VLA & 4.8 & 12.9 & $117 \pm 18$ \\
May 13.15 & VLA & 7.4 & 12.9 & $166 \pm 16$ \\
May 14.03 & VLA & 21.8 & 28.5 & $133 \pm 23$ \\
May 23.96 & VLA & 4.8 & 8.7 & $191 \pm 25$ \\
May 23.96 & VLA & 7.4 & 8.7 & $248 \pm 18$ \\
June 1.06 & VLA & 22.5 & 26.5 & $118 \pm 20$ \\
June 12.88 & VLA & 4.8 & 13.8 & $140 \pm 26$ \\
June 12.88 & VLA & 7.8 & 13.8 & $117 \pm 17$ \\
June 26.86 & VLA & 4.8 & 14.0 & $158 \pm 19$ \\
June 26.86 & VLA & 7.4 & 14.0 & $160 \pm 15$ \\
July 9.94 & VLA & 4.8 & 14.0 & $62 \pm 23$ \\
July 9.94 & VLA & 7.4 & 14.0 & $98 \pm 21$ \\
Aug. 21.71 & VLA & 8.5 & 37.2 & $< 44$ \\
Dec. 16.45 & VLA & 8.4 & 25.0 & $63 \pm 20$ \\ 
\enddata
\label{tab:radio}
\end{deluxetable}


\clearpage
\begin{deluxetable}{lcc}
\tabletypesize{\scriptsize}
\tablecaption{PTF GRB Simulation Results}
\tablewidth{0pt}
\tablehead{
\colhead{$N$} & \colhead{$N_{\mathrm{Det}}$} &
\colhead{$N_{\mathrm{Obs}}$}
}
\startdata
  1 & 1311 & 11376 \\
  2 & 1583 & 40101 \\
  3 & 228 & 5889 \\
  4 & 118 & 825 \\
  5 & 30 & 693 \\
  6 & 31 & 305 \\
  7 & 6 & 189 \\
  8 & 4 & 113 \\
  9 & 3 & 54 \\
  $> 10$ & 26 & 426 \\
\enddata
\label{tab:Ndet}
\end{deluxetable}

\end{document}